# A ferroelectric junction transistor memory made from switchable van der Waals p-n heterojunctions

Baoyu Wang[1,2,3#], Lingrui Zou[1,2,4#], Tao Wang[1,2], Lijun Xu[5], Zexin Dong[6], Chuan Qin[2], Xin He[1,2], Shangui Lan[1,2], Yinchang Ma[3], Meng Tang[3], Maolin Chen[3], Chen Liu[3], Zheng-Dong Luo[7,8], Lijie Zhang[4]*, Zhenhua Wu[1], Yan Liu[7,8], Genquan Han[7,8], Bin Yu[2], Xixiang Zhang[3]*, Kai Chang[1]*, and Fei Xue[1,2]*

[1] Center for Quantum Matter, School of Physics, Zhejiang University, Hangzhou 310027, China

[2] College of Integrated Circuits, ZJU-Hangzhou Global Scientific and Technological Innovation Center, Zhejiang University, Hangzhou 311215, China

[3] Physical Science and Engineering Division, King Abdullah University of Science and Technology, Thuwal 23955-6900, Saudi Arabia

[4] Key Laboratory of Carbon Materials of Zhejiang Province, College of Chemistry and Materials Engineering, Wenzhou University, Wenzhou 325035, Zhejiang, China

[5] Institute of Microelectronics, Chinese Academy of Sciences, Beijing 100029, China

[6] School of Semiconductor Science and Technology, South China Normal University, Foshan 528225 China

[7] Hangzhou Institute of Technology, Xidian University, Hangzhou 311200, China

[8] State Key Discipline Laboratory of Wide BandGap Semiconductor Technology, School of Microelectronics, Xidian University, Xi'an 710071, China

* Corresponding author: Lijie Zhang, ljzhang@wzu.edu.cn; Xixiang Zhang, xixiang.zhang@kaust.edu.sa; Kai Chang, kchang@zju.edu.cn; Fei Xue, xuef@zju.edu.cn

[#] These authors contributed equally.

**Van der Waals (vdW) p-n heterojunctions are important building blocks for advanced electronics and optoelectronics, in which high-quality heterojunctions essentially determine device performances or functionalities. Creating tunable depletion regions with substantially suppressed leakage currents presents huge challenges, but is crucial for heterojunction applications. Here, by using band-aligned p-type SnSe and n-type ferroelectric $\alpha$-$In_2Se_3$ as a model, we report near-ideal multifunctional vdW p-n heterojunctions with small reverse leakage currents (0.1 pA) and a desired diode ideality factor (1.95). We realize ferroelectric-tuned band alignment with a giant barrier modulation of 900 meV. Based on such tunable heterojunctions, we propose and demonstrate a fundamental different memory device termed ferroelectric junction field-effect transistor memory, which shows large memory windows (1.8 V), ultrafast speed (100 ns), high operation temperature (393 K), and low cycle-to-cycle variation (2%). Additionally, the reliable synaptic characteristics of these memory devices promise low-power neuromorphic computing. Our work provides a new device platform with switchable memory heterojunctions, applicable to high performance brain-inspired electronics and optoelectronics.**

**Teaser sentence:** Switchable van der Waals p-n heterojunctions enable a novel ferroelectric junction field-effect transistor (Fe-JFET) for low-power neuromorphic electronics.

## INTRODUCTION

Heterostructures integrate the unique advantages of two distinct materials and expand the functionalities beyond single-material systems, which lay the foundation for modern electronic technology. In particular, two-dimensional (2D) van der Waals (vdW) heterostructures, which are assembled through weak interlayer forces without the constraints of lattice matching, exhibit remarkable versatility compared to conventional silicon-based heterostructures. The atomically smooth and dangling-bond-free surfaces effectively eliminate interfacial defects, granting exceptional 2D heterostructure properties, such as ultra-thin thickness, strong interlayer coupling, tailorable band alignment and efficient charge transfer (*1-5*). These properties have greatly propelled extensive 2D heterostructure research into creating a variety of post-Moore electronic and optoelectronic devices, including diodes (*6-8*), photodetectors (*9-13*), light-emitting diodes (*14-16*), tunnel field-effect transistors (*17-19*), and junction field-effect transistors (JFETs) (*20-24*).

Currently, ferroelectric material-based 2D heterostructures have attained growing interest because polarization offers a new knob to tune physical properties and craft high-performance devices (*25-36*). However, due to the imperfect depletion region, large leakage currents across ferroelectric polarized heterostructures can heavily deteriorate device performance and cause increased power consumption overhead. To mitigate this issue, constructing high-quality heterostructures with fully depleted regions and electrically tunable barriers is crucial. Here, we determine the specific band alignment of heterostructures by selecting a p-type SnSe and n-type ferroelectric semiconductor α-$In_2Se_3$. Thanks to the ultraclean and flat vdW interfaces, we realize high-quality p-n heterojunctions and introduce ferroelectric polarization charges over the depletion region. By using these heterojunctions as an example, we show an unprecedented device concept and its design, which we term ferroelectric junction field-effect transistor memory (Fe-JFET). We envision that our heterojunction design provides a versatile platform for developing multifunctional, high-performance JFETs that incorporate information storage, photo-sensing, light-emitting, and energy-harvesting functionalities.

Specifically, the vdW p-n heterojunction demonstrates a high rectifying ratio over $2\times10^4$, and an ideality factor of 1.95 comparable to the ideal diode (*6, 37, 38*). The Fe-JFET shows excellent transistor performance including a high on/off ratio over $10^5$, and low subthreshold swing (SS) down to 138 mV/dec. In addition, the Fe-JFET exhibits competitive nonvolatile memory behaviors, such as a large memory window (1.8 V), ultrafast switching (100 ns), high operation temperature (393 K), and low cycle-to-cycle variation (2%). Piezoresponse force microscopy (PFM) and Kelvin probe force microscopy (KPFM) reveal that the underlying principle of memory behaviors is attributed to ferroelectric-tuned band alignment. Furthermore, we simulated a convolutional neural network based on the Fe-JFET, highlighting its potential for neuromorphic computing applications.

## RESULTS

### The vdW Fe-JFET structure

Conventional ferroelectric field-effect transistors (Fe-FET) are predominantly based on metal-insulator-semiconductor (MIS) structure and can be categorized into two types depending on the role of ferroelectrics (Fig. 1A). As shown in the Fe-FET I type (left panel of Fig. 1A), the structure incorporates a ferroelectric dielectric (ferro-dielectric) as the gate insulator layer (*39-42*). The resistance hysteresis is modulated by the carrier accumulation or depletion from ferroelectric polarization charges over the channel. By contrast, for the Fe-FET II type (right panel of Fig. 1A), a ferroelectric semiconductor is used as the channel (ferro-channel), rather than the gate dielectric layer (*43-46*). In this configuration, the ferro-channel polarization can be modulated by gate voltage, with polarization charges influencing band bending and enabling multiple resistance states. Although both Fe-FETs' configurations have demonstrated excellent performance, the dielectric layer inevitably introduces charge trapping at the interface between dielectric layer and semiconductor channel, resulting in threshold voltage drift, current reduction and poor polarization fatigue (*47-49*). Consequently, further efforts in interface engineering are required to minimize these adverse impacts (*50, 51*).

To potentially sidestep these limitations, we propose a fundamental different Fe-JFET device without an active dielectric layer (Fig. 1B). In this configuration, an n-type narrow-bandgap ferroelectric semiconductor is used as the ferro-channel to leverage its switchable polarization and excellent electrical properties. Moreover, a p-type semiconductor, with a higher doping concentration than the channel, is utilized as the gate to not only minimize the voltage drops on the gate but also enable complete junction depletion (*52*). This mechanism is fundamentally distinct from that of conventional FeFETs, which rely on polarization-induced carrier accumulation or depletion in the ferroelectric layer. A detailed mechanism is elaborated in the following sections.

### Heterojunction and materials characterization

Fig. 2A shows the atomic force microscopy (AFM) morphology of Fe-JFET, where exfoliated multilayer n-type α-$In_2Se_3$ and p-type SnSe flakes were sequentially transferred onto Si substrates. SnSe layer with a thickness of approximately 60 nm, and α-$In_2Se_3$ layer exhibiting strong ferroelectricity and excellent semiconducting properties at typical thicknesses of 20–50 nm were adopted, as illustrated in fig. S1.

Back-gate transfer characteristic of p-type SnSe is shown in fig. S2, which indicates channel's incomplete depletion and high carrier concentration. When fabricating the device, metal stacks (Cr/Au) were deposited on both ends of SnSe and α-$In_2Se_3$ flakes and Ohmic contacts were intentionally formed at drain and source terminals to achieve excellent transistor performance.

To verify the crystal phase and quality, we used various techniques of material characterization. Fig. 2B presents typical Raman spectra of separate SnSe and *α*-$In_2Se_3$ flakes. The peak positions for *α*-$In_2Se_3$ crystal are consistent with the previous report (*53*), indicating the hexagonal phase. X-ray diffraction (XRD) characterization was performed, as shown in fig. S3. The distinct *c*-plane peaks of α-$In_2Se_3$ demonstrate a well-defined layered hexagonal structure, which belongs to the $P6_3mc$ space group (NO. 186)(*46*). Photoluminescence measurement was also conducted on both *α*-$In_2Se_3$ and p-n junction region (Fig. 2C). The intense peak at 885 nm from *α*-$In_2Se_3$ suggests a direct bandgap of 1.40 eV. In contrast, the junction region shows a significantly weaker and red-shifted peak, which can be attributed to incident light absorption by SnSe and carrier recombination within the junction. It is worth noting that SnSe with an indirect bandgap material would not emit a distinct photoluminescence peak (*54*).

For exploring α-$In_2Se_3$ ferroelectric polarization, we performed second-harmonic generation (SHG) and PFM measurements. Fig. S4 shows a strong SHG signal excited by 1064 nm pulse laser, reflecting the intrinsic noncentrosymmetry. Polar plots of the SHG signals (Fig. 2D) reveal six-fold asymmetric property, which is the origin of the presence of ferroelectric polarization (*55*). To examine the ferroelectric switching of α-$In_2Se_3$ crystal, we wrote a square pattern (Fig. 2E) by applying +8 V onto the PFM tip. The results exhibit distinct phase change with inward polarization in the inner region. Additionally, a standard off-field amplitude butterfly hysteresis with 180° phase reversal shown in Fig. 2F further demonstrates the reversal of ferroelectric polarization.

Subsequent structural characterization and analysis were performed on the as-fabricated Fe-JFET device. The cross-sectional scanning transmission electron microscope (STEM) image (Fig. 2G) clearly displays a high-resolution interface for the p-n heterojunction. Correlated energy dispersive spectroscopy (EDS) analysis identifies the exact elements within each individual layer. The right panel of Fig. 2G shows atomic crystal structures of the orthorhombic SnSe and hexagonal α-$In_2Se_3$ lattice, coinciding with previous studies (*55, 56*). Owing to the van der Waals interface, there is no lattice mismatch in the p-n junction region. The clean atomically interfaces ensure the high quality of the formed p-n heterojunction, which is vital for exceptional Fe-JFET performance.

The depletion quality of the p-n heterojunction in Fe-JFETs plays a crucial role in determining device performance. We first plotted the band alignment of the heterojunction (fig. S5), in which the 1 eV and 1.40 eV bandgaps of SnSe and $\alpha$-$In_2Se_3$ were adopted, respectively. The energy band offsets in the spontaneous state are approximately 0.61 eV ($\Delta E_C$) and 1.01 eV ($\Delta E_V$), which dominate the electrical rectification upon small voltage sweeping. Next, we investigated the p-n diode characteristics using the two-terminal architecture shown in fig. S6A, wherein the bottom $\alpha$-$In_2Se_3$ source is grounded and the external bias is applied through the top SnSe gate. In this configuration, the vertical rectifying current ($I_{Rec}$) is driven across the heterojunction interface. The measured I-V curve shown in Fig. 2H demonstrates excellent rectifying behavior with an exceptionally low reverse saturation current below 0.1 pA and the threshold voltage of 0.6 V. Correspondingly, the bias-dependent band diagrams in fig. S6B elucidate the substantial modulation of the junction energy barrier, which fundamentally dictates the consequent forward carrier injection and reverse carrier depletion. To quantitatively evaluate the rectifying performance, ideality factor ($n$) was estimated at a small forward bias (here is 0.3-0.7 V) by fitting to the below diode equation (*6*):

$$I = I_S[\exp\left(\frac{V}{nV_T}\right)-1]$$

where $I$, $I_S$, $V$ and $V_T$ denote diode current, reverse saturation current, applied bias voltage, and thermal voltage, respectively. The fitted $n$ is approximately 1.95, comparable to the previously reported near-ideal p-n and Schottky junction diodes (*6, 37, 38*). Ultimately, such robust diode rectification indicates a high-quality junction (*20, 52, 57*), enabling the low leakage current, small threshold voltage, and high on/off ratio in the Fe-JFET.

**Fe-JFET electrical characterization**

First, the transistor performance of the Fe-JFET was characterized by applying the gate voltage to SnSe ($V_{GS}$) and the source-drain voltage ($V_{DS}$) to the $\alpha$-$In_2Se_3$ ferro-channel. Fig. 3A presents typical transfer curves at $V_{DS}$ = 1 V showing a high on/off ratio over $10^5$, which surpasses that of most $\alpha$-$In_2Se_3$ FETs based on metal-dielectric-semiconductor structures (*44-46, 58-61*). The gray dashed curves indicate ultra-low leakage currents down to 1 pA with respect to $V_{GS}$, and the variation is attributed to the p-n hetero-junction transition between conduction and depletion states. Note that, for both ferroelectric-narrowed or enlarged depletion regimes (forward and backward sweeping), such ultra-low leakage currents are commonly observed in our fabricated Fe-JFETs, as shown in fig. S7, which is important for our transistor memories. Moreover, Ohmic contact is essential for effective channel current injection and extraction, thereby reducing heating loss and enhancing switching speed in transistors. We characterized Ohmic contact through the high linearity as in Fig. 3B, which shows the output characteristics at various $V_{GS}$ under $V_{DS}$ sweeping from -0.25 to 0.25 V. As the gate voltage varies from -3 V to 2 V, the channel current is effectively modulated from off transistor state to on transistor state. In Fig. 3C, the $V_{DS}$ sweeping range is extended to 2 V for further characterizing the output characteristics of the Fe-JFET. The operational regions transitioning from suppression state to linear and then to saturation regions are clearly observed, which is dependent on the applied $V_{GS}$ and $V_{DS}$ values, indicating effective control over the channel current.

After investigating the remarkable transistor performance, we explored the unique memory characteristics imparted by the α-$In_2Se_3$ ferro-channel in the Fe-JFET. As shown in Fig. 3D, the transfer curves, measured by bidirectional $V_{GS}$ sweep within a small range from -3 V to 2 V at various $V_{DS}$, exhibit a distinctive clockwise hysteresis. This large hysteresis, resulting from the ferroelectric polarization switching in the ferro-channel, reflects the transition of non-volatile high and low resistance states. Despite the smooth vdW interface between α-$In_2Se_3$ and SnSe, we conducted comparative experiments to rule out the possibility that the transfer hysteresis is caused by trapping effects. We selected $MoS_2$, which is without ferroelectricity, as the channel to fabricate JFETs and measured their transfer characteristics under the same conditions, as shown in fig. S8. Consistent with previous reports (*52, 57, 62*), no hysteresis behavior is observed, implying that the transfer hysteresis essentially originates from ferroelectric polarization reversal over the α-$In_2Se_3$ ferro-channel.

With different fixed $V_{DS}$ values, the transfer curves upon varying $V_{GS}$ sweeping ranges are shown in fig. S9. The hysteresis becomes larger as the $V_{GS}$ range is expanded, indicating the gradual flipping of ferro-channel polarization. Fig. 3E presents the hysteresis transconductance ($g_m$-$V_{GS}$) calculated from the transfer hysteresis at different $V_{DS}$. The maximum $g_m$ reached 0.8 μS and 0.6 μS under forward and reverse $V_{GS}$ sweep, respectively, demonstrating the excellent gate control capability of our Fe-JFET. Notably, the Fe-JFET achieves an SS of as low as 138 mV/dec (Fig. 3F), which is, to the best of our knowledge, significantly superior to that of Fe-FET II type based on metal-dielectric-semiconductor structures (*43, 46, 58-61*). The performance of Fe-JFET at both low (fig. S10) and high temperatures (fig. S11) were also thoroughly investigated, as described in Supplementary Text S1. Even at a high temperature of 393 K, the Fe-JFET can display obvious resistance switching.

We turn to evaluate the memory performance of our proposed Fe-JFET. Compared to conventional JFET, the ferroelectric polarization from α-$In_2Se_3$ ferro-channel and the high-quality vdW heterojunctions endow Fe-JFET with the unique functionalities of memory and ultrafast writing that are absent in previously reported JFETs (*20-24, 52, 57, 62-65*). Fig. 3G shows the modulation of the Fe-JFET's current state by a voltage pulse with 100 ns width, resulting in a significant current decrease. The current switching with respect to different pulse widths is shown in fig. S12. Notably, the degree of resistance variation is directly dependent on the pulse width. Moreover, Fig. 3H and 3I present the endurance and retention characteristics of the Fe-JFET. After 1500 cycles of writing (+4 V, 100 ms) and erasing (-4 V, 100 ms) operations, negligible degradation is detected and both high/low resistance states persist for at least 1000s. A slight decrease in high/low resistance states may result from the depolarization field and potential read-disturb effects during continuous monitoring. However, despite this decay, a highly distinguishable on/off ratio is maintained throughout the measurement period, ensuring reliable state readout for memory applications. In addition, the $V_{GS}$ waveform and $I_{DS}$ variation measured during the endurance test are presented in fig. S13. After each positive or negative pulsed $V_{GS}$ poling, the $I_{DS}$ measured at the read voltage ($V_{read}$) exhibits a high or low state, respectively. Compared with documented Fe-FETs (including type I and II) and JFETs, our Fe-JFET demonstrates excellent performance in both transistor and memory functionalities. For detailed comparison, published relevant works have been summarized in the Supplementary Table.

**Fe-JFET working mechanism**

Having established the concept of Fe-JFET and characterized the electrical performance, we next probed the underlying mechanism that enables the transistor and memory functionalities. We conducted an in-depth analysis through in-situ PFM and KPFM measurements and verified by density functional theory (DFT) calculation. Following previous works(*52, 57, 65*), the physical explanation for transistor functionality can be found in Supplementary Text S2, where transfer curve analysis and Technology Computer Aided Design (TCAD) simulations (fig. S14) are provided, and the pinch-off voltage is also extracted (fig. S15). For memory mechanism, we have speculated that the resistance hysteresis arises from ferroelectric polarization switching at the heterojunction. To explore this dynamics, firstly the three-terminal Fe-JFET architecture was evaluated (Fig. 4A), comprising the gate on SnSe and the $\alpha$-$In_2Se_3$ channel contacted by source and drain electrodes. The operating principle is captured in the transfer characteristics at $V_{DS}$ = 1 V (Fig. 4B), where a pronounced hysteresis loop defines two distinct resistance states (on and off states) governed by ferroelectric polarization switching. To corroborate this polarization-driven mechanism, in-situ spatial characterizations were performed on the device. Following gate poling at -4 V and +4 V, PFM phase mappings of the channel (Fig. 4C) exhibit a stark contrast, confirming a robust reversal of the ferroelectric polarization between outward and inward states. Subsequently, KPFM was utilized to map the exact barrier modulations at the p-n heterojunction (Fig. 4D). The specific potential difference was extracted from the red and blue curves marked in the two poled states and was re-plotted in Fig. 4E.

Because KPFM measurement only probes surface potentials and p-type SnSe flake is too thick, no distinct potential variation is detected in heterojunction region but a large variation of approximately 900 meV is seen on $\alpha$-$In_2Se_3$ surface. This potential change can be used to assess the junction barrier variation. We note that such a giant barrier modulation warrants high-performance memory devices, such as large memory windows and high on/off ratios.

Driven by these electrical and in-situ characterizations, we formulate the corresponding energy band diagrams for the on and off states. As shown in Fig. 4F, the redistributed positive and negative polarization charges dynamically tune the energy band alignment at the p-n heterojunction interface. Specifically, in the on state, the upward polarization ($P_{up}$) induces positive bound charges at the heterointerface, which narrows the depletion width ($W_D$) and lowers the energy barrier, thereby facilitating carrier transport. Conversely, in the off state, the downward polarization ($P_{down}$) accumulates negative bound charges, substantially widening the depletion region and elevating the barrier to suppress current flow. Correspondingly, the bottom panels of Fig. 4F illustrate the band alignment along the channel length, where the localized positive and negative polarization charges induce downward and upward band bending, respectively. In this way, the Fe-JFET can be effectively switched between the on and off states. The polarization-modulated band alignment is further substantiated by DFT calculations. As shown in fig. S16, the projected band structures reveal that switching from the on to the off state triggers a massive relative shift between the α-$In_2Se_3$ (red) and SnSe (blue) energy bands, theoretically corroborating the large barrier variations at the heterojunction. Overall, the dynamic modulation of the heterojunction barrier via ferroelectric polarization switching constitutes the fundamental operating principle of the Fe-JFET.

**Neuromorphic computing with Fe-JFET**

Hardware capable of executing neuromorphic computing holds great potential in addressing the substantial computational demands brought by the advent of the artificial intelligence (AI) era (*66, 67*). Leveraging the excellent performance of the Fe-JFET, we explored its potential applications in brain-inspired neuromorphic computing. Towards this goal, we attempted to fundamentally mimic the biological behaviors of synapses including long-term and short-term plasticity, and spike-dependent synaptic plasticity. The SnSe gate electrode can function as a presynaptic terminal and the α-$In_2Se_3$ ferro-channel represents postsynaptic current (PSC). Pulsed $V_{GS}$ emulate biological spikes, while the variations of channel current correspond to synaptic weight updates. Typical learning rule, such as long-term plasticity, was emulated by applying 40 consecutive pulses of -2 V, 4 ms and +1 V, 4 ms onto $V_{GS}$, as shown in Fig. 5A, with a total of 10 cycles performed. Benefiting from the robust non-volatile ferroelectric polarization, the

α-$In_2Se_3$ channel current was progressive excitatory and inhibitory, corresponding to synaptic long-term potentiation (LTP) and long-term depression (LTD) characteristics.

Cycle-to-cycle variation (CCV), which represents the fluctuation in each conductance state across multiple cycles, reflects the synaptic device's repeatability and robustness (*68*). A small CCV can mitigate the degradation in learning accuracy typically caused by high nonlinearity, and hence perform better in neuromorphic computing. The CCV can be defined as:

$$\mathrm{CCV}_{\,(\mathrm{n,\,j})} = \frac{\left| G_{\mathrm{n}}^{j+1} - G_{\mathrm{n}}^{j} \right|}{G_{\max} - G_{\min}} \times 100\%$$

where $G_n$ represents the conductance state, $j$ represents the cycle index and $G_{max}$ and $G_{min}$ are the maximum and minimum conductance values, respectively. As shown in Fig. 5B, the CCV over 10 cycles is predominantly below 2%, indicating excellent neuromorphic characteristics of the Fe-JFET. Additionally, two alternative pulse schemes were implemented to simulate the complex biological stimulation signals (fig. S17), with a detailed discussion provided in Supplementary Text S3.

Modulating the amplitude of the stimulation pulses from -0.5 to -2 V in the LTP process results in a notable increased PSC (Fig. 5C). This indicates the expanded dynamic range of Fe-JFET and its ability to emulate spike-dependent synaptic plasticity. Paired-pulse facilitation (PPF) is a form of short-term synaptic plasticity where two closely timed stimuli result in an enhanced PSC response to the second stimulus (*69*). By applying a pair of consecutive pulse $V_{GS}$ to the Fe-JFET and varying the time interval ($\Delta$t) between the pulses, the PPF index was obtained (Fig. 5D). The magnitude is higher at short intervals and the facilitation gradually declines as the Δt increases, following a biexponential decay function. This behavior is consistent with the behavior of biological synapses. Subsequently, sequential negative and positive $V_{GS}$ pulses were applied to the Fe-JFET to explore dynamic LTP and LTD characteristics. Fig. 5E depicts that the PSC was initially potentiated to a high level by varying 40 negative $V_{GS}$ pulses and then suppressed to the initial state by applying 40 positive $V_{GS}$, demonstrating flexible synaptic plasticity.

Furthermore, building on the above synaptic performance of the Fe-JFET, we developed a convolutional neural network (CNN) model to simulate system-level image recognition. As shown in Fig. 5F, the 28×28-pixel MNIST (Modified National Institute of Standards and Technology) handwritten digit dataset was used as training data, with our Fe-JFET acting as the synaptic connections between neurons. During the backpropagation learning algorithm, the weight update rules rely on the experimentally measured LTP and LTD characteristics of the Fe-JFET under sequential $V_{GS}$ pulses. Further details of CNN deep learning are provided in Supplementary Text S3. To quantitatively evaluate the image recognition capability of the constructed CNN model, 10000 test images were classified. The confusion matrix presented in Fig. 5G demonstrates a high recognition accuracy across various classes. After 50 training epochs, with each epoch utilizing 50000 training images, the model achieved an accuracy of 95.8%, as shown in Fig. 5H. The step-by-step image feature extraction process is shown in fig. S18. These results highlight the potential of Fe-JFET-based CNN models in efficient image recognition applications, suggesting promising avenues for future research in neuromorphic computing and system-level pattern recognition tasks. Further optimization of Fe-JFET synaptic properties and exploration of more complex datasets could expand the applicability of this approach, advancing the integration of ferroelectric devices in next-generation AI hardware.

## DISCUSSION

We have reported on a new memory concept of Fe-JFET and shown its design and mechanism by using switchable SnSe/$\alpha$-$In_2Se_3$ p-n heterojunction with suppressed leakage currents and a desired diode ideality factor. The Fe-JFET memory devices exhibit a large memory window (1.8 V), ultrafast operation speed (100 ns), and high operation temperature (393 K), and low cycle-to-cycle variation (2%). Furthermore, we showed the synaptic behaviors of the Fe-JFET and constructed a CNN model for image recognition with high accuracy.

The Fe-JFET represents an unprecedented device platform with ferroelectric-tuned depletion regions. The memory heterojunctions used in this device are not restricted to conductive gate and ferroelectric semiconducting channel, but can be extended to many other band-aligned ferroelectric stackings, such as ferroelectric tellurium gate/black phosphorus channel or sliding ferroelectric BN gate/$MoS_2$ semiconductor channel. This prototype device would have broad applications in self-polarized, nonvolatile, high-performance electronics or optoelectronics, such as solar cells, LEDs, and photodetectors, and in their low-power neuromorphic computing with all-in-one functionalities.

## MATERIALS AND METHODS

### Device fabrication

Multilayer $\alpha$-$In_2Se_3$ and SnSe were obtained by mechanical exfoliation from bulk crystals. Then they were sequentially transferred onto heavily doped Si substrate through a 2D materials transfer stage. The gate, source and drain electrodes were defined by laser direct writing (Heidelberg DWL66+) and electron beam lithography (Crestec CABL 9000C), followed by the deposition of 40 nm Cr and 10 nm Au via electron-beam evaporation (Denton Vacuum Explorer). Finally, the photoresist was lifted off by sequentially soaking in acetone and isopropanol solutions.

## Materials characterization

*Raman and SHG measurements*

Raman measurements were conducted using the WITec alpha300 apyron confocal Raman spectroscopy. The excitation light source was a 532 nm continuous-wave laser. A 100×/NA 0.95 Zeiss objective lens was used to focus the laser onto the sample. An 1800 grooves/mm grating was employed for spectral dispersion during the measurements. SHG measurements were also performed using WITec alpha300 apyron using the same test conditions but the excitation light source was a 1064 nm picosecond pulsed laser.

*XRD measurements*

The single-crystal XRD pattern was measured using the Bruker D8 Discover X-ray diffractometer with CuKα radiation. The system was operated in θ-2θ mode to analyze the crystallographic structure of the sample. The measurements were performed with a generator voltage of 40 kV and a tube current of 40 mA to ensure adequate X-ray intensity.

*PL measurements*

PL was measured by using Horiba LabRAM Odyssey. A 100× VIS objective lens was used to focus the laser onto the sample, and a 100 grooves/mm grating was employed for spectral dispersion. A neutral density filter with 10% transmission was applied to control the laser intensity.

*PFM and KPFM measurements*

Asylum Research MFP-3D and Bruker Dimension Icon were used to characterize the device morphology, as well as for PFM and KPFM measurements. In-situ PFM measurement was performed in dual ac resonance tracking (DART) mode with an SCM-PIT-V2 type conductive tip (spring constant of 3 N/m) at a resonance frequency of approximately 280 kHz to acquire amplified piezoelectric responses. In-situ KPFM measurement was conducted in AM-KPFM mode on the Bruker Dimension Icon using the same conductive probe with a tip lift height set to 30 nm. The Fe-JFET device was mounted on a PCB board and connected to the Keithley 2400 source meter via wire bonding, which provided the gate voltage for both in-situ PFM and KPFM measurements.

## Electrical measurements

The transistor electrical performance of the Fe-JFET was measured using the Keithley 4200, and memory and synaptic properties were evaluated with the Agilent B1500. During the measurement, the two gate electrodes at the ends of the SnSe layer were shorted to obtain enhanced gate control over α-$In_2Se_3$ ferro-channel. All measurements were conducted on a homemade probe station under ambient conditions at room temperature. Temperature-dependent electrical measurements were also conducted on the homemade probe station. The chamber was evacuated to a low vacuum of $5\times10^{-2}$ Torr, with temperature control achieved by heating with a hot plate and cooling provided by liquid nitrogen.

### TCAD simulation and DFT calculation

The Sentaurus TCAD simulator was employed to simulate the proposed Fe-JFET device, ensuring an accurate analysis of its working mechanism of transistor functionality. The device structure was carefully designed using the Sentaurus Structure Editor (Sde), where the size parameters of device modeling were consistent with the experiments. The $\alpha$-$In_2Se_3$ channel featured source and drain Ohmic contacts at its two ends, and the top SnSe layer was defined as the gate electrode. The bandgaps of $\alpha$-$In_2Se_3$ and SnSe are set to 1.40 eV and 0.9 eV, respectively. Physical models including mobility, recombination, quantum potential and barrier tunneling were employed in the simulation.

All DFT computational results were obtained using the commercial software QuantumATK (version 2023.12), which employs the linear combination of atomic orbitals (LCAO) method. The calculations were performed based on the Generalized Gradient Approximation (GGA) with the Perdew-Burke-Ernzerhof (PBE) exchange-correlation function. The interaction between valence electrons and ionic nuclei was described using PseudoDojo pseudopotentials. The DFT-1/2 method was applied to correct the bandgap of materials.

### CNN simulation

The CNN simulation for MNIST digit recognition and classification was implemented using Python in PyCharm IDE. The process involved defining the CNN model architecture, which includes convolutional, activation, pooling, and fully connected layers, followed by training and testing the network on standard datasets. The trained CNN obtained optimal weight parameters, with the weight updates during backpropagation adapted to strictly follow the LTP and LTD characteristics of the Fe-JFET device through a revised SGD algorithm. Finally, the network's performance was evaluated for its recognition accuracy, demonstrating the potential of Fe-JFET-based neural networks in classification tasks. Detailed information about weight updates can be found in Supplementary Text S3.

## Acknowledgements

B. -Y. W. acknowledges the support from the China Scholarship Council (CSC, No. 202306320423).

**Funding:** This research was supported by the National Key R&D Program of China under grant No. 2023YFB4402303 (Z.-D. L.), the National Science Foundation of China (grant No. 62304202 to F. X., No. 62274128 to Z.-D. L., No. 92264202 to Z.-D. L., No. 62090033 to Y. L.), the Zhejiang Provincial Natural Science Foundation of China under grant No. LDT23F0402 (Z.-D. L.), LDT23F04023F04 (Z.-D. L.) and No. LDT23F04024F04 (G. -Q. H.), and the Key Research and Development Program of Ningbo City under No. 2023Z071 (Z.-D. L.). This research was also supported by the Fundamental Research Funds for the Central Universities (No. K20250157 to F. X.).

**Author contributions:** F. X. conceived the research. B. -Y. W., L. -R. Z., T. W., M. T., and M. -L. C. fabricated devices and tested their electrical characteristics. B. -Y. W., L. -R. Z., T.W., C. L., and Y. -C. M. performed the materials and device characterization. B. -Y. W., L.-J. X., Z. -X. D., C. Q., and Z. -H. W. performed simulations and calculations. F. X., B. -Y. W., X. H., and S. -G. L. analyzed the data. B. -Y. W. and F. X. wrote the paper. Z.-D. L., Y.L., and G. -Q. H. improved the paper. F. X., B. Y., X.-X. Z., L. -J. Z., and K. C. supervised this project.

**Competing interests:** F. X., B. -Y. W., and B. Y. are inventors listed on the Chinese patent ZL202411472536.3 (granted date: 11 December 2024), held by ZJU-Hangzhou Global Scientific and Technological Innovation Center, that covers a ferroelectric JFET device, its fabrication method, and applications. All other authors declare no competing interests.

**Data and materials availability:** All data needed to evaluate the conclusions in the paper are present in the paper and/or the Supplementary Materials.

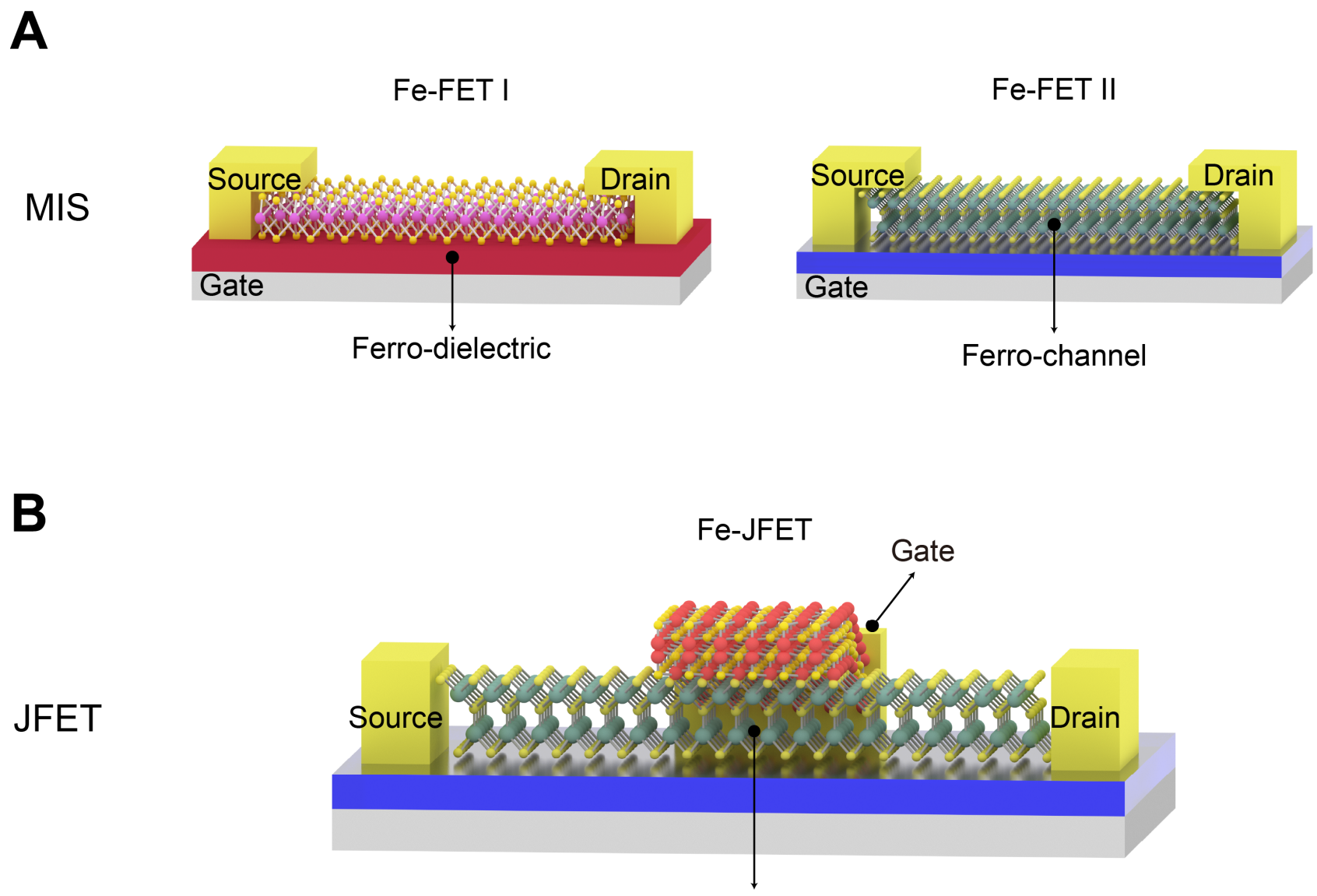

**Fig. 1. Device architectures of conventional Fe-FETs and the proposed Fe-JFET. (A)** Illustration of two types of conventional ferroelectric field-effect transistors (Fe-FET I and II), where either a ferroelectric gate dielectric (Ferro-dielectric) or a ferroelectric channel (Ferro-channel) is used. The polarization directions, e.g., $P_{down}$ and $P_{up}$, influence the distribution of mobile carriers over the channel, resulting in resistance switching. **(B)** The proposed novel structure that we term a ferroelectric junction field-effect transistor (Fe-JFET) memory, which combines a p-type semiconductor with an n-type ferroelectric channel (Ferro-channel). Source, drain and gate terminals are marked.

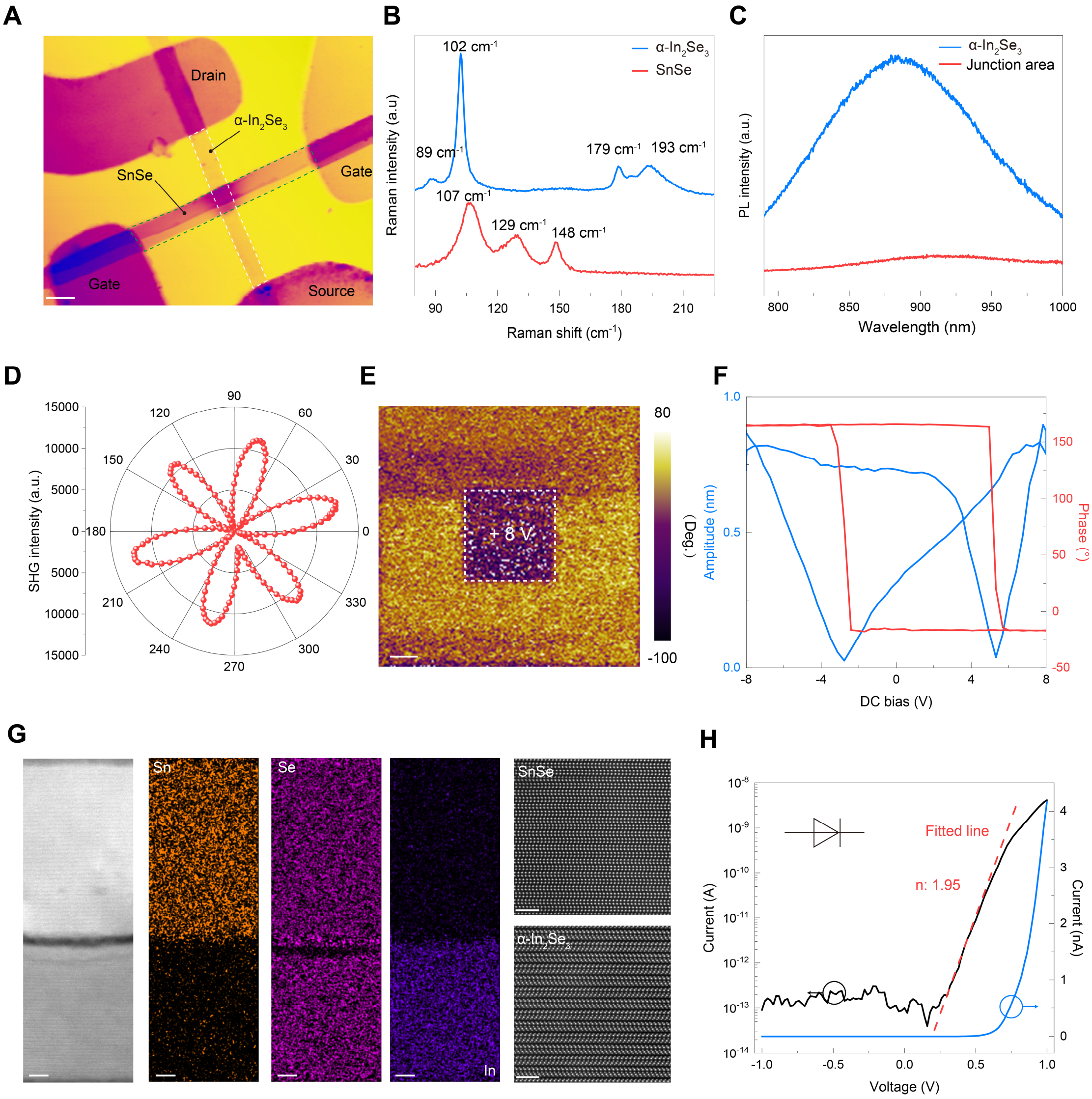

**Fig. 2. Characterization of SnSe/α-$In_2Se_3$ heterojunctions. (A)** AFM morphology of the SnSe/α-$In_2Se_3$ Fe-JFET structure. Scale bar: 500 nm. **(B)** Raman spectra of α-$In_2Se_3$ and SnSe flakes. The characteristic peaks confirm the crystal phase. **(C)** Photoluminescence spectra of the α-$In_2Se_3$ and junction area. **(D)** SHG intensity of α-$In_2Se_3$, indicating its non-centrosymmetric ferroelectric nature. **(E)** Box-written PFM phase mapping of ferroelectric α-$In_2Se_3$ shows ferroelectric polarization switching. Scale bar: 400 nm. **(F)** PFM amplitude and phase responses as a function of DC biases, demonstrating the local switching of ferroelectric polarization. **(G)** Cross-sectional TEM images of the SnSe/α-$In_2Se_3$ heterojunction show the layered structure and vdW interface. Elemental mappings for Sn, Se, and In confirm the composition of the heterojunction. Scale bar: 5 nm. Atomic crystal structures of the orthorhombic SnSe and hexagonal α-$In_2Se_3$. Scale bar: 2 nm. **(H)** Rectification characteristic of the vdW SnSe/α-$In_2Se_3$ p-n heterojunction.

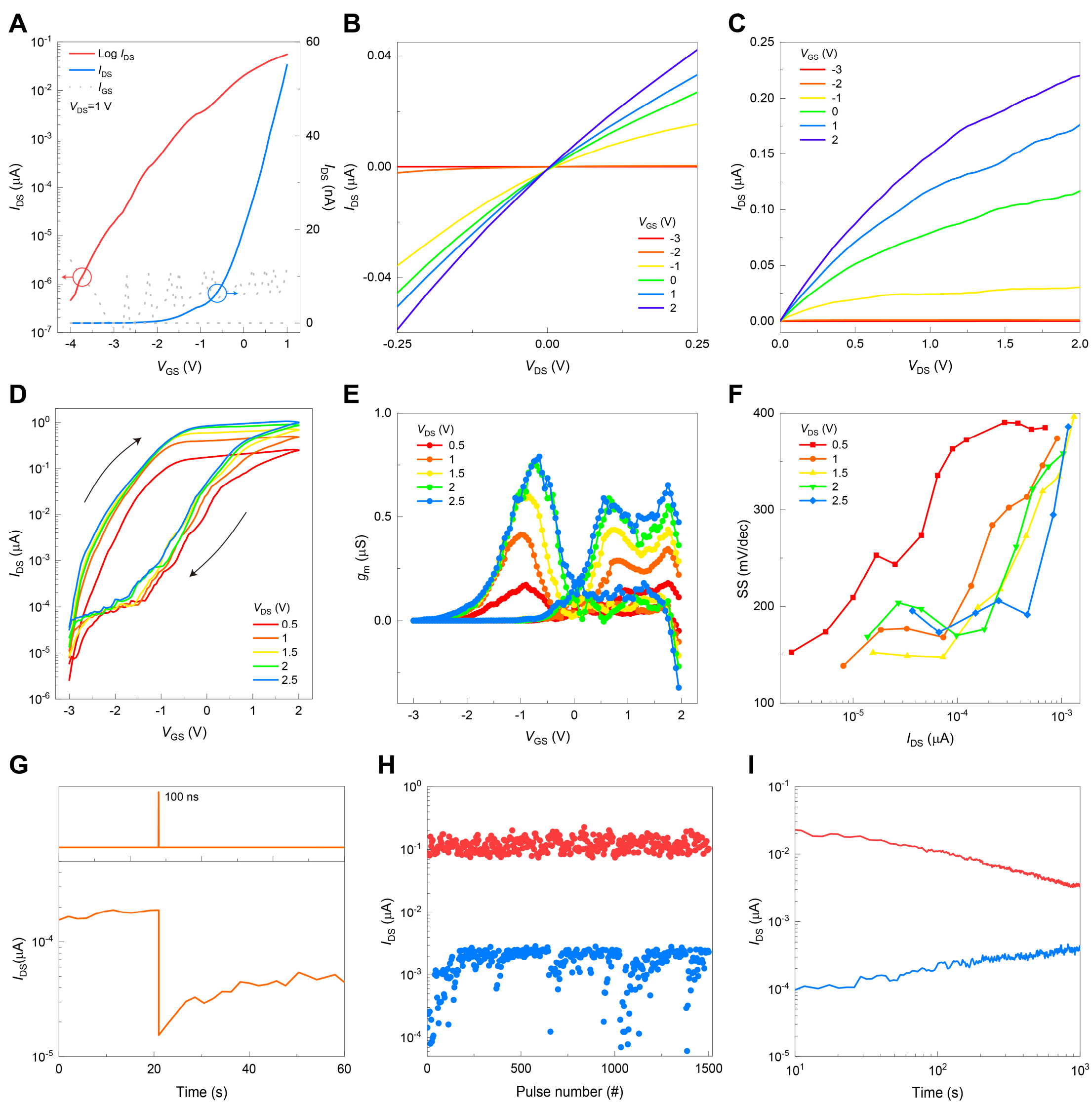

**Fig. 3. Electrical characterization of the SnSe/α-$In_2Se_3$ Fe-JFET. (A)** Transfer curves at $V_{DS}$ =1 V. Grey curves dictate leakage currents across the heterojunction. **(B)** Output curves under different $V_{GS}$ from -3 to 2V. The high linearity demonstrates Ohmic contact. **(C)** Output curves over an extended $V_{DS}$ range with varying $V_{GS}$, clearly illustrating the distinct operating regions. **(D)** Bidirectional scanning transfer curves display a large memory window, which originates from the polarization switching of α-$In_2Se_3$ ferro-channel. **(E)** Transconductance ($g_m$) as a function of $V_{GS}$ at various $V_{DS}$ from 0.5 to 2.5V. **(F)** SS versus $I_{DS}$ for different $V_{DS}$. The raw data in **E** and **F** is taken from **D**. **(G)** Dynamic current change simulated by a short voltage pulse (10 V, 100 ns). The Fe-JFET exhibits ultra-fast memory operation. **(H)** Endurance tests over 1500 cycles show both stable high and low resistance states. The $I_{DS}$ was measured at $V_{DS}$ = 1 V, $V_{GS}$ = -1 V. **(I)** Retention test for both two resistance states at $V_{DS}$ = 1 V and $V_{GS}$ = -1 V.

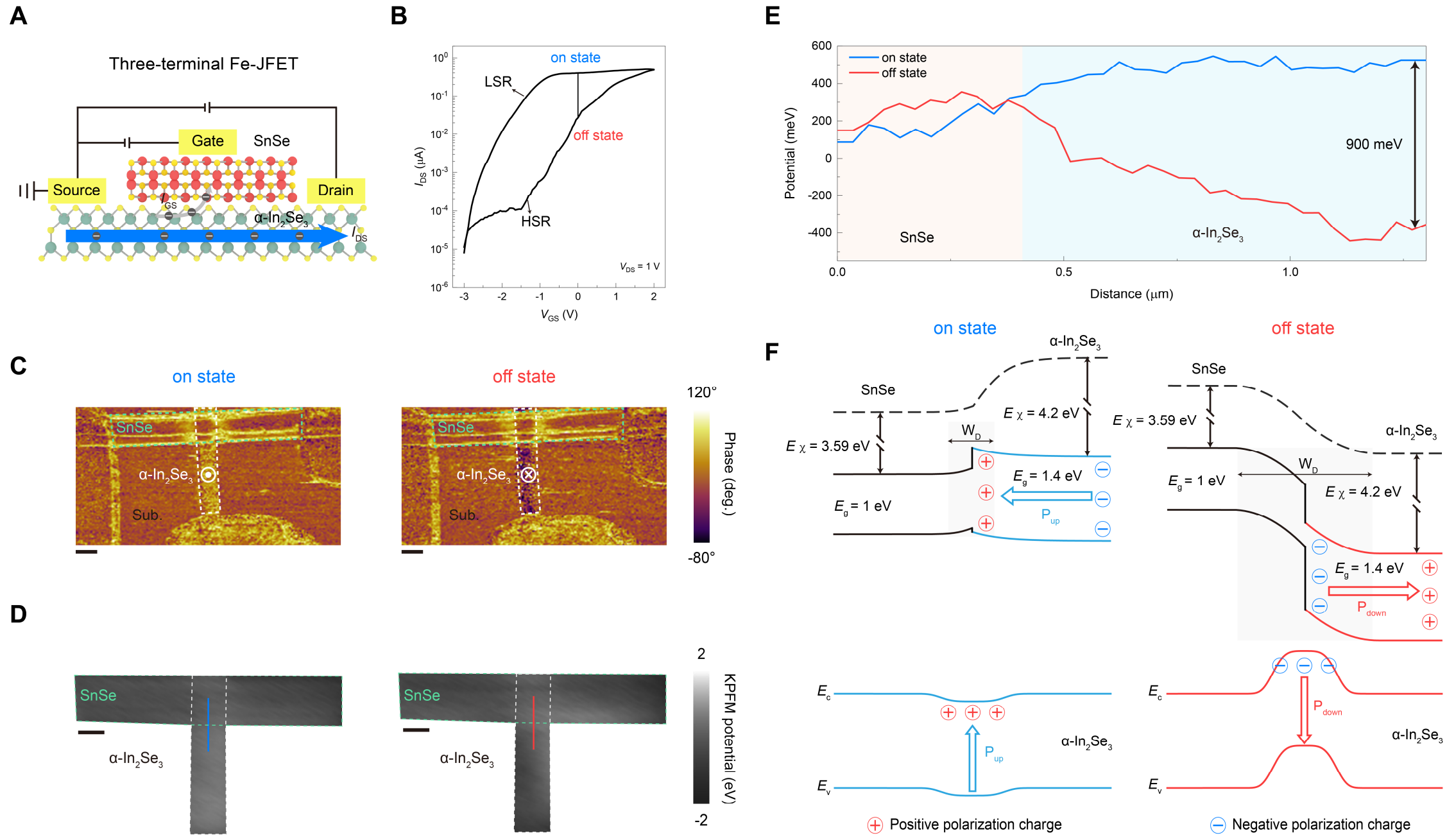

**Fig. 4. Fe-JFET working mechanism. (A)** Schematic diagram of the connection configuration for the device working in three-terminal mode as a Fe-JFET. **(B)** Transfer curve of the Fe-JFET measured at $V_{DS}$ = 1 V. The explicit hysteresis loop clearly defines the low-resistance state (LSR, on state) and high-resistance state (HSR, off state). **(C)** PFM phase mapping of the α-$In_2Se_3$ ferro-channel after -4 V and +4 V poling at $V_{GS}$. Scale bar: 500 nm. **(D)** KPFM mapping of the surface potential at the SnSe/α-$In_2Se_3$ p-n junction after poling, illustrating potential variations caused by ferroelectric polarization switching. Scale bar: 500 nm. **(E)** Specific potential differences extracted from the given red and blue curves in the two poled states, showing a barrier variation of approximately 900 meV. **(F)** Energy band diagrams of the SnSe/α-$In_2Se_3$ p-n heterojunction in the on and off states. The switchable ferroelectric polarization ($P_{up}$ and $P_{down}$) of the α-$In_2Se_3$ channel drives a non-volatile variation in the depletion width ($W_D$) and energy barrier, determining the distinct states of the Fe-JFET.

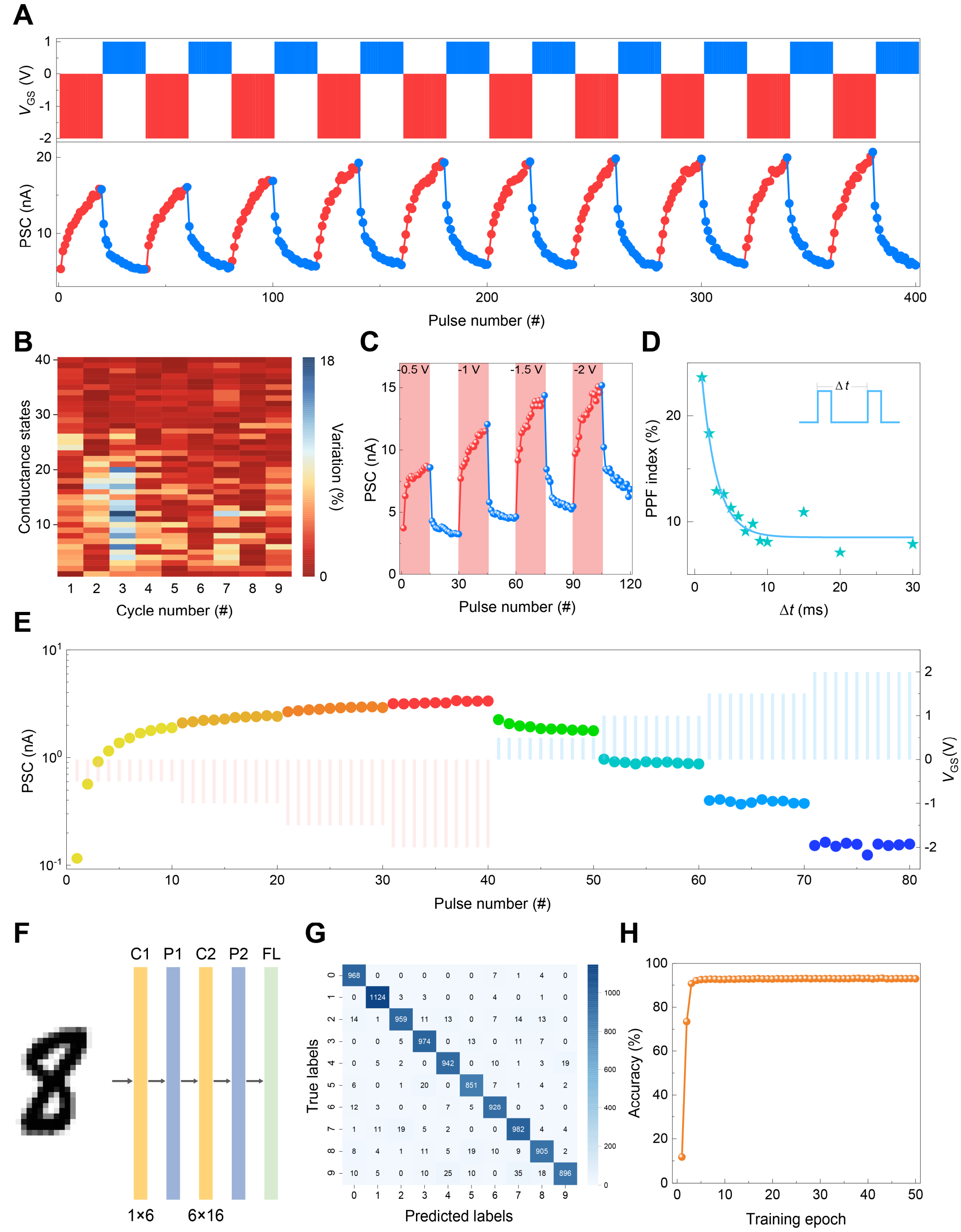

**Fig. 5. Synaptic behaviors of Fe-JFET for neuromorphic computing. (A)** Postsynaptic current (PSC) responses to pulsed $V_{GS}$, exhibiting LTP and LTD characteristics over 10 pulse cycles. Top panel is the pulse voltage scheme and bottom panel is the corresponding PSC response. **(B)** CCV heatmap indicates the stability of PSC across a multitude of cycles. **(C)** Measurement of amplitude-dependent plasticity. PSC potentiation increases with higher $V_{GS}$ amplitudes from -0.5 to -2V. **(D)** Extracted PPF index as a function of the time interval between two pulsed $V_{GS}$. The solid line is fitted with a double exponential function. **(E)** Dynamic PSC potentiation and depression behaviors under several sequential pulsed $V_{GS}$ with various amplitudes. **(F)** Schematic of the CNN model with the Fe-JFET acting as synaptic weights. **(G)** Confusion matrix for image recognition of the MNIST handwritten digit dataset. **(H)** Image recognition accuracy of the CNN model over 50 training epochs, reaching a steady accuracy of 95.8%.

Supplementary Materials for

# A ferroelectric junction transistor memory made from switchable van der Waals p-n heterojunctions

Baoyu Wang *et al*.

Corresponding author: Lijie Zhang, ljzhang@wzu.edu.cn; Xixiang Zhang, xixiang.zhang@kaust.edu.sa; Kai Chang, kchang@zju.edu.cn; Fei Xue, xuef@zju.edu.cn

**This PDF file includes:**

Supplementary Texts S1 to S3

Figs. S1 to S18

Supplementary Table

**Supplementary Text S1: Temperature-dependent electrical performance.**

To further investigate the electrical performance of as-fabricated Fe-JFET, we conducted low temperature-dependent electrical measurements. The transfer loops at 77 K and 100 K are shown in fig. S10A and S10B, respectively. It can be seen that the current was decreased, which is consistent with previous works (*70, 71*). While the hysteresis loop became more pronounced compared to the room-temperature counterparts. This phenomenon is mainly attributed to the reduced intrinsic carrier concentration ($n_i$), which is exponentially dependent on temperature, as described by equation (S1), leading to the decreased current at cryogenic temperature.

$$n_{\mathrm{i}} \propto \sqrt{N_{\mathrm{C}} N_{\mathrm{V}}} \cdot \mathrm{e}^{-\frac{E_{\mathrm{g}}}{2k_{\mathrm{B}}\mathrm{T}}} \quad \text{(S1)}$$

where $N_C$ and $N_V$ are the effective density of states in the conduction and valence bands, $E_g$ is the bandgap energy, $k_B$ is Boltzmann's constant and $T$ is temperature. Low temperatures reduce the thermal activation of carriers in the channel and suppress the trapping effect at the interface, resulting in improved gate control capability and a more pronounced hysteresis loop. Although phonon scattering is reduced and electron mobility increases at low temperature, the transconductance decreases correspondingly due to the reduced carrier concentration, weakened thermal activation effect, and the resulting decrease in channel current (fig. S10C and S10D). SS is fundamentally limited by thermal energy, as described by the below equation:

$$\mathrm{SS} = \ln 10 \cdot \frac{k_{\mathrm{B}}\mathrm{T}}{\mathrm{q}} \cdot \left(1 + \frac{C_{\mathrm{dep}}}{C_{\mathrm{ox}}}\right) \quad \text{(S2)}$$

At lower temperatures, the thermal energy term is reduced carrier mobility is increased (*72*), leading to a significant reduction in the SS, as shown in fig. S10E-S10H.

We further assessed the temperature tolerance of our Fe-JFET by evaluating its performance under continuously increasing temperatures, ranging from 303 K to 423 K. As shown in fig. S11, the hysteresis window induced by ferroelectric polarization switching gradually diminishes with increasing temperature and vanishes entirely at 423 K. However, when the temperature returns to room temperature, a noticeable hysteresis window reappears, likely due to partial recovery of ferroelectric polarization following thermal depolarization at elevated temperatures.

**Supplementary Table: Overview for comparing vdW material-based Fe-FETs and JFETs.**

| Structure | Gate | Channel | On/off ratio | SS (mV/dec) | Memory & MW/Sweep range | Switching time & Temperature |
|---|---|---|---|---|---|---|
| MIS (*42*)<br>Ferro-dielectric | CIPS | $MoS_2$ | $10^7$ | < 60 | No | N. A |
| MIS (*40*)<br>Ferro-dielectric | HZO | $MoS_2$ | $10^6$ | < 60 | No | N. A |
| MIS (*41*)<br>Ferro-dielectric | P(VDF-TrFE) | $MoTe_2$ | $10^4$ | Large | No | N. A |
| MIS (*73*)<br>Ferro-dielectric | AlScN | $MoS_2$ | $10^7$ | Large | Yes<br>18 V/ 30 V | N. A |
| MIS (*74*)<br>Ferro-dielectric | MOF | $MoS_2$ | $10^7$ | Large | Yes<br>8.3 V/ 70 V | N. A |
| MIS (*45*)<br>Ferro-dielectric | BN/$\alpha$-$In_2Se_3$ | $MoS_2$ | $10^5$ | Large | Yes<br>14 V/ 26 V | N. A |
| MIS (*51*)<br>Ferro-dielectric/<br>Trapping | CIPS/BN | $MoS_2$ | $10^7$ | 17 | Yes<br>3.8 V/ 8 V | 100 ns<br>N. A |
| MIS (*46*)<br>Ferro-channel | h-BN | $\alpha$-$In_2Se_3$ | $10^4$ | Large | Yes<br>6 V/ 16 V | 40 ns<br>423 K |
| MIS (*75*)<br>Ferro-channel | $Al_2O_3$ | $\alpha$-$In_2Se_3$ | $10^5$ | Large | Yes<br>12.4 V/ 20 V | N. A |
| MIS (*44*)<br>Ferro-channel | $SiO_2$ | $\alpha$-$In_2Se_3$ | $10^4$ | Large | Yes<br>24.1 V/ 80 V | N. A |
| MIS (*76*)<br>Ferro-channel | $SiO_2$ | $\gamma$-$In_2Se_3$ | $10^7$ | Large | Yes<br>40.3 V/ 80 V | N. A |
| MIS (*77*)<br>Ferro-channel | $Al_2O_3$ | $\beta$-$In_2Se_3$ | $10^7$ | Large | Yes<br>2 V/ 15 V | N. A<br>423 K |
| MIS (*78*)<br>Ferro-channel | BN | 3R $MoS_2$<br>(Sliding) | $10^6$ | Large | Yes<br>7 V/ 10 V | 1 μs<br>N. A |
| JFET (*57*) | $PdSe_2$ | $MoS_2$ | $10^4$ | 150 | No | N. A |
| JFET (*52*) | SnSe | $MoS_2$ | $10^6$ | 60 | No | N. A |
| JFET (*79*) | $MoTe_2$ | $MoS_2$ | $10^4$ | 100 | No | N. A |
| JFET (*64*) | $NiO_X$ | $MoS_2$ | $10^5$ | 60 | No | N. A |
| JFET (*80*) | BP | ZnO | $10^3$ | 200 | No | N. A |
| JFET (*62*) | $NbS_2$ | $MoS_2$ | $10^6$ | 67 | No | N. A |
| **Fe-JFET<br>Ferro-channel<br>(our work)** | **SnSe** | **$\alpha$-$In_2Se_3$** | **$10^5$** | **138** | **Yes<br>1.8 V/ 5 V** | **100 ns<br>393 K** |

**Supplementary Text S2: Operating principle of the transistor functionality.**

We analyzed the working mechanism of the Fe-JFET transistor functionality based on its electrical characteristics and validated it using TCAD simulations. Fig. S14A depicts the linear and logarithmic plots of $I_{DS}$, along with the corresponding $I_{GS}$. It is evident that the $I_{GS}$ increases significantly at reverse bias -3.74 V, which we define as $V^{-}_{Leaky}$. The pinch-off voltage $V_P$ of -1.54 V is estimated from the $\sqrt{I_{DS}}$versus $V_{GS}$ curve (fig. S15). Note that the channel current $I_{DS}$ originates from lateral electron transport along the α-$In_2Se_3$ channel driven by $V_{DS}$, while the leakage current $I_{GS}$ arises from the vertical current across the SnSe/α-$In_2Se_3$ p-n junction, which includes diffusion current under forward bias and drift current under reverse bias. Specifically, there is a competitive relationship between the lateral channel current $I_{DS}$ and the vertical leakage current $I_{GS}$. When $V^{-}_{Leak}<V_{GS}<V_P$, both the p-n junction and the transistor channel are in the cut-off state, resulting in minimal $I_{DS}$ and $I_{GS}$ (formed by drift current in the p-n heterojunction). When $V_{GS}< V^{-}_{Leak}$, the p-n junction undergoes weak reverse breakdown, causing a significant rise in $I_{GS}$ due to the breakdown current (fig. S14B top). When $V_{GS}>V_P$, the p-n junction slightly opens, and the channel also transitions conductive state. $I_{GS}$ formed by diffusion current remains much smaller than the $I_{DS}$ because the p-n junction is incompletely forward-biased (fig. S14B bottom). Specially, as $V_{GS}$ continues to increase, the diffusion current through the fully forward conduction p-n junction grows exponentially, eventually causing $I_{GS}$ to exceed the saturated $I_{DS}$.

To better understand how the p-n junction states determine the transistor functionality, a TCAD simulation was performed to analyze the electric field distribution across the junction under various $V_{GS}$. As shown in fig. S14C, when $V_{GS} < V_P$, the reverse biased p-n junction sustains a large voltage drop due to the extensive depletion region. At $V_{GS} = V_P$, the electric field decreases as the junction begins to open, enabling partial charge transport. When $V_{GS} > V_P$, the electric field further reduces, indicating that the forward-biased junction is fully open, enabling strong carrier flow across the channel. These results reflect the dependence of the electric field on the $V_{GS}$, also demonstrating the effective modulation of the p-n junction by $V_{GS}$.

**Supplementary Text S3: The CNN simulation.**

To demonstrate the synaptic plasticity of the Fe-JFET under complex stimulation signals, its LTP/LTD characteristics were tested using two distinct pulse schemes: one with gradually increased pulse width (pulse scheme2) and the other with gradually increased pulse amplitude (pulse scheme3). As shown in fig. S17A and S17B, the Fe-JFET exhibits relatively high linearity under these two pulse schemes. The extracted maximum and minimum conductance values ($G_{max}$ and $G_{min}$) from the LTP/LTD characteristics, as shown in fig. S17C, remain consistently stable within a defined range. The CCV heatmaps over 10 cycles for the two pulse schemes 2 and 3 shown in fig. S17D and S17E, respectively, exhibit a small variation, further demonstrating the excellent robustness of our Fe-JFET as a neuromorphic synaptic device. Next, a representative example of LTP and LTD from each of the three pulse schemes was selected for a more detailed performance analysis, as shown in fig. S17F-S17H. The nonlinear factors for LTP range from 0.8 to 2.0, while those for LTD range from 1.5 to 3.3, indicating a relatively high degree of symmetry. The asymmetry ($\beta_{asy}$, defined as the absolute difference between the nonlinear factors of LTP and LTD) reaches a minimum value of 9.144, providing a solid foundation for subsequent neural network computations (*75*).

The CNN simulation was implemented in Python within the PyCharm IDE. The CNN model consists of two convolutional layers (C1, C2) followed by activation functions, two pooling layers (P1, P2), and a fully connected layer (FL) with three layers. The final output layer comprises 10 neurons for digits 0-9 recognition. C1 has 6 output channels with a kernel size of 5, stride of 1, and padding of 2, followed by a Sigmoid activation and P1 with a kernel size of 2 and stride of 2. C2 has 16 output channels, also with a kernel size of 5 and stride of 1, but without padding, followed by Sigmoid activation and P2 with the same parameters. Following the convolutional and pooling layers, a flattening layer prepares the feature map for the FL. The fully connected (linear) layers include three layers with decreasing dimensions: 400 to 120, 120 to 84, and finally 84 to 10, where the output layer consists of 10 neurons for classification purposes. The forward function sequentially passes the input through each layer, resulting in a structured flow from feature extraction to classification.

$$G_{n+1} = G_n + \Delta G_P = G_n + \alpha_P e^{\beta_P (G_n - G_{min} / G_{max} - G_{min})} \quad \text{(S3)}$$

$$G_{n+1} = G_n + \Delta G_D = G_n - \alpha_D e^{\beta_D (G_{max} - G_n / G_{max} - G_{min})} \quad \text{(S4)}$$

We utilized above two equations(*68*), where $\Delta G$ represents the variation of synaptic weight, $G_{n+1}$ and $G_n$ denote synaptic weight after $n+1_{th}$ and $n_{th}$ modulation respectively to fit the LTP/LTD characteristics of our Fe-JFET. The obtained parameters of $\alpha_P$, $\beta_P$, $\alpha_D$, and $\beta_D$ were used to define a strict update rule for synaptic weights governed by equations (S3) and (S4). This update rule was implemented in an adapted stochastic gradient descent (SGD) algorithm for weight updates during backpropagation, ensuring that the weight adjustments strictly adhered to the LTP and LTD characteristics of our Fe-JFET.

Using the digit “8” as the input image, we further illustrate the recognition process with the well-trained CNN model. As shown in fig. S18, the input image is processed through two convolutional layers, each followed by sigmoid activation and average pooling, extracting essential features at each stage. In the output layer, the model predicts “8” with a high confidence of 98.18 %.

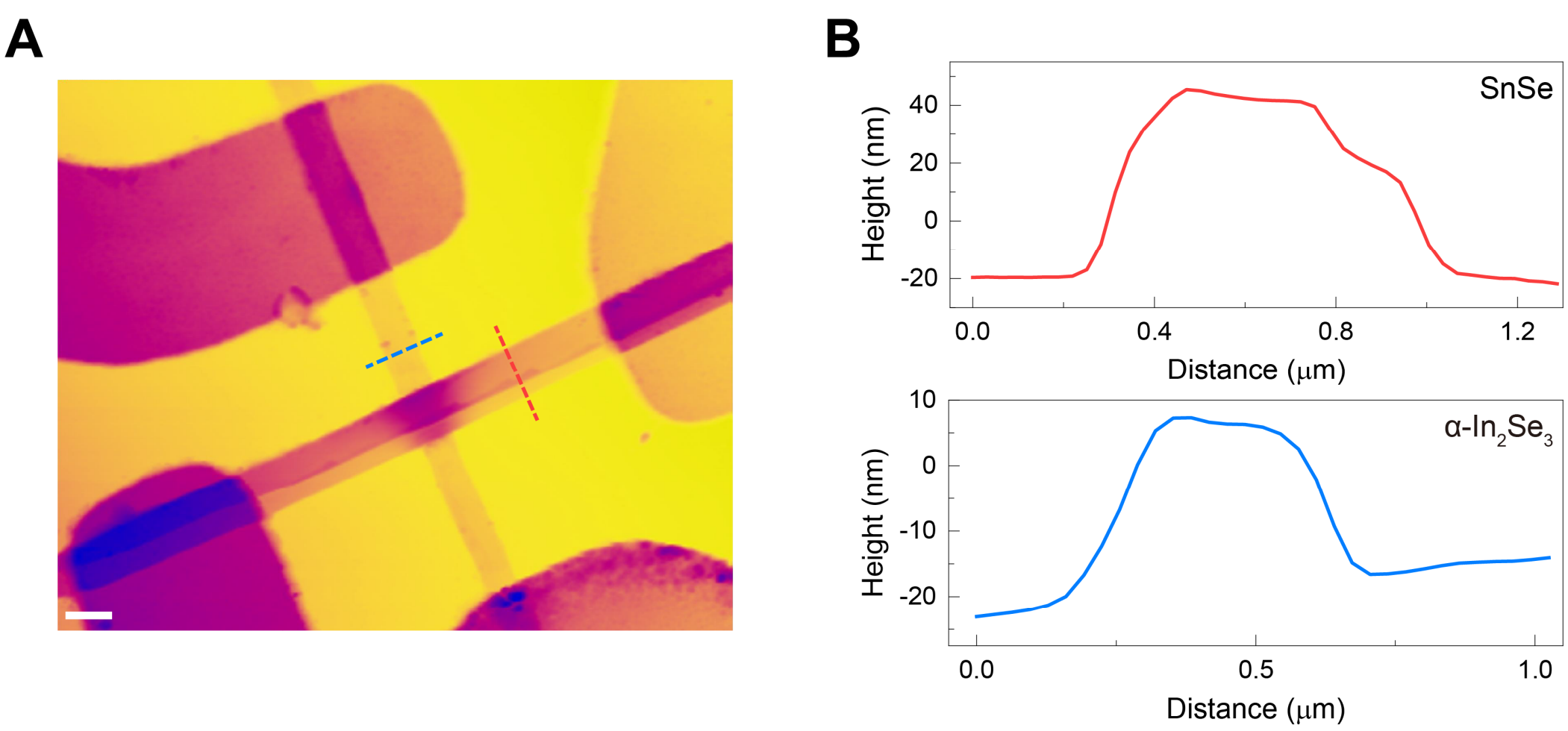

**Fig. S1. Morphological characterization of the fabricated Fe-JFET. (A)** AFM topography image. Scale bar: 500 nm. **(B)** Thicknesses of SnSe and α-$In_2Se_3$ layers. They are approximately 60 nm and 30 nm, respectively.

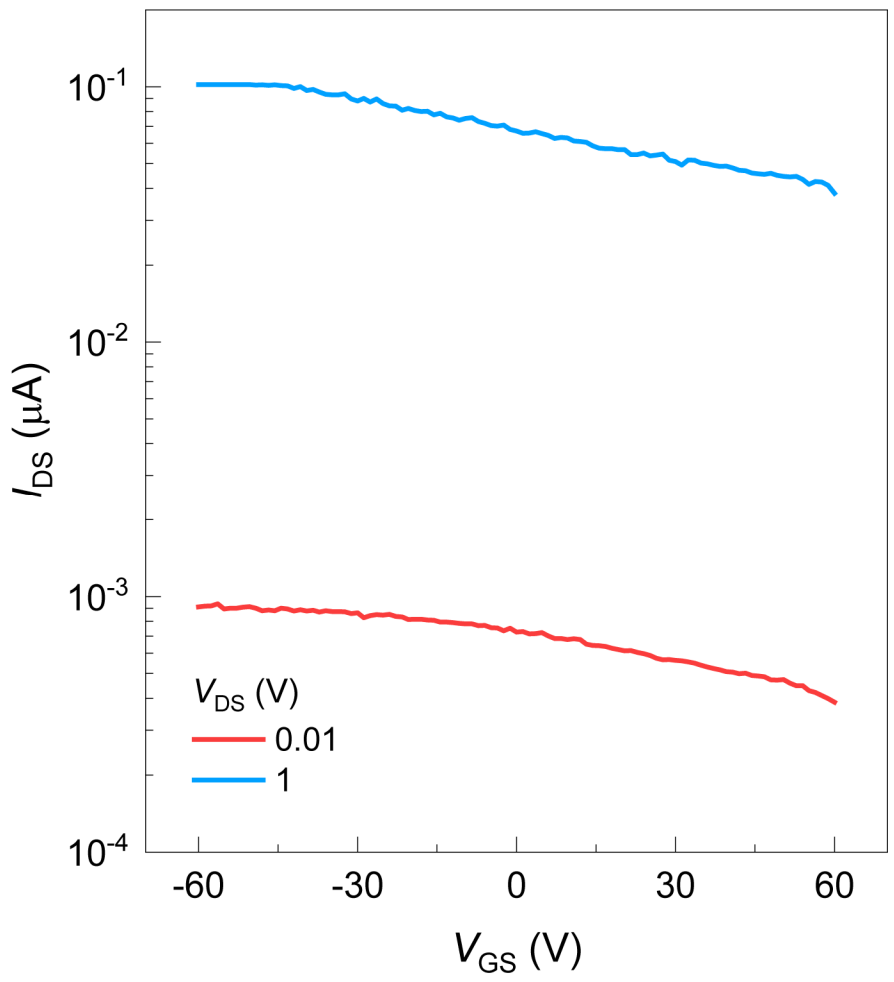

**Fig. S2. Back gate transfer characteristics of p-type SnSe.** A small back-gate tunability indicates incomplete depletion and high carrier concentration.

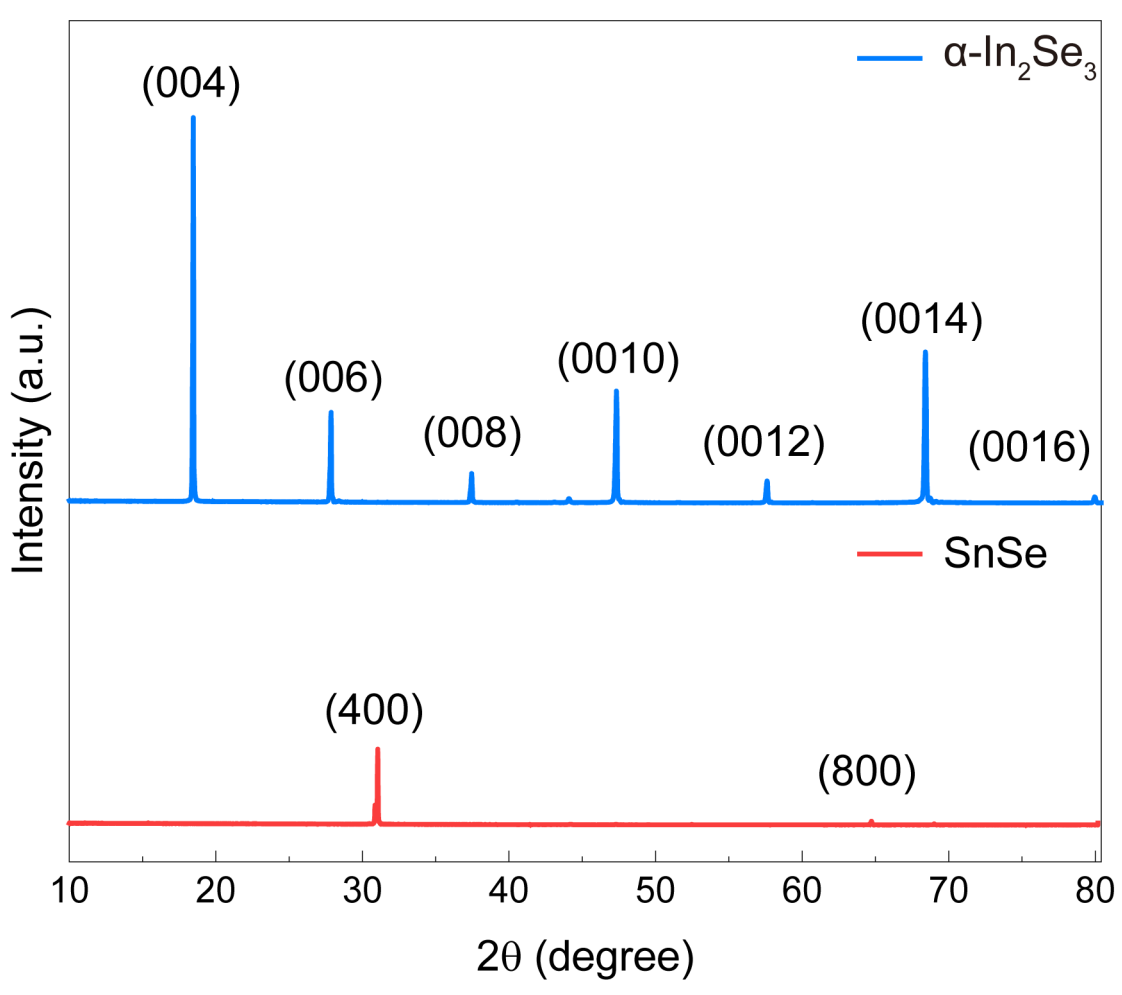

**Fig. S3. XRD characterization of α-$In_2Se_3$ and SnSe.**

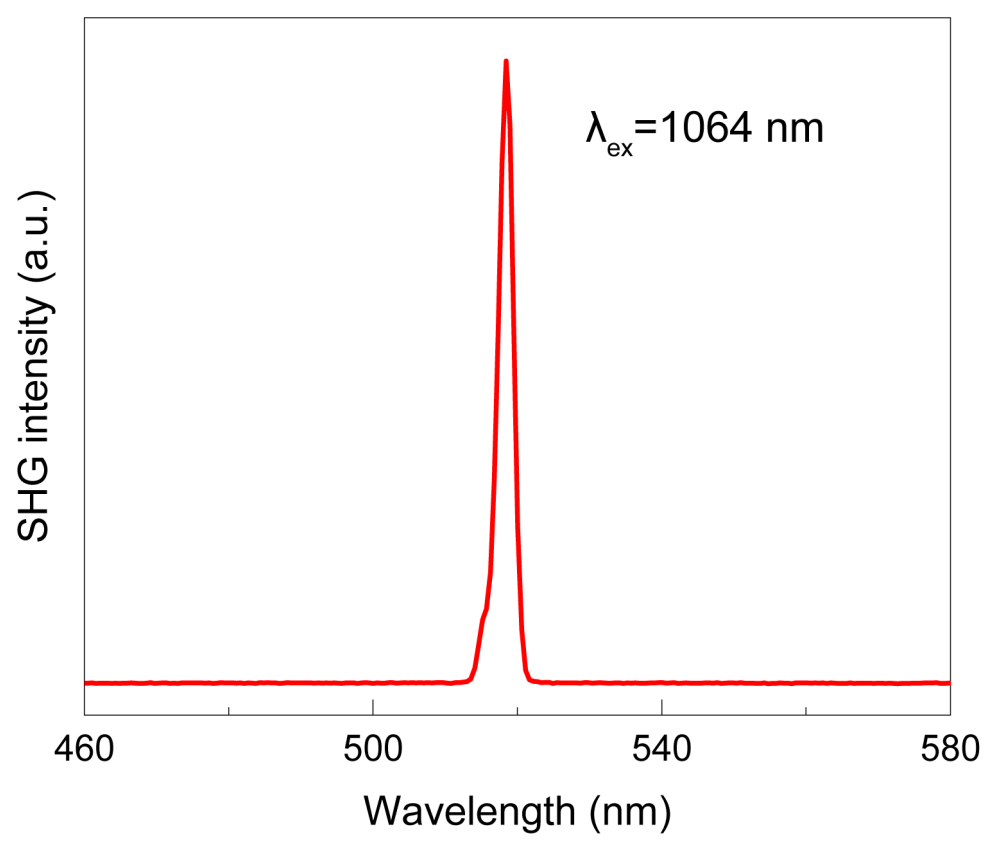

**Fig. S4. Strong SHG signal of α-$In_2Se_3$ excited by 1064 nm pulse laser.**

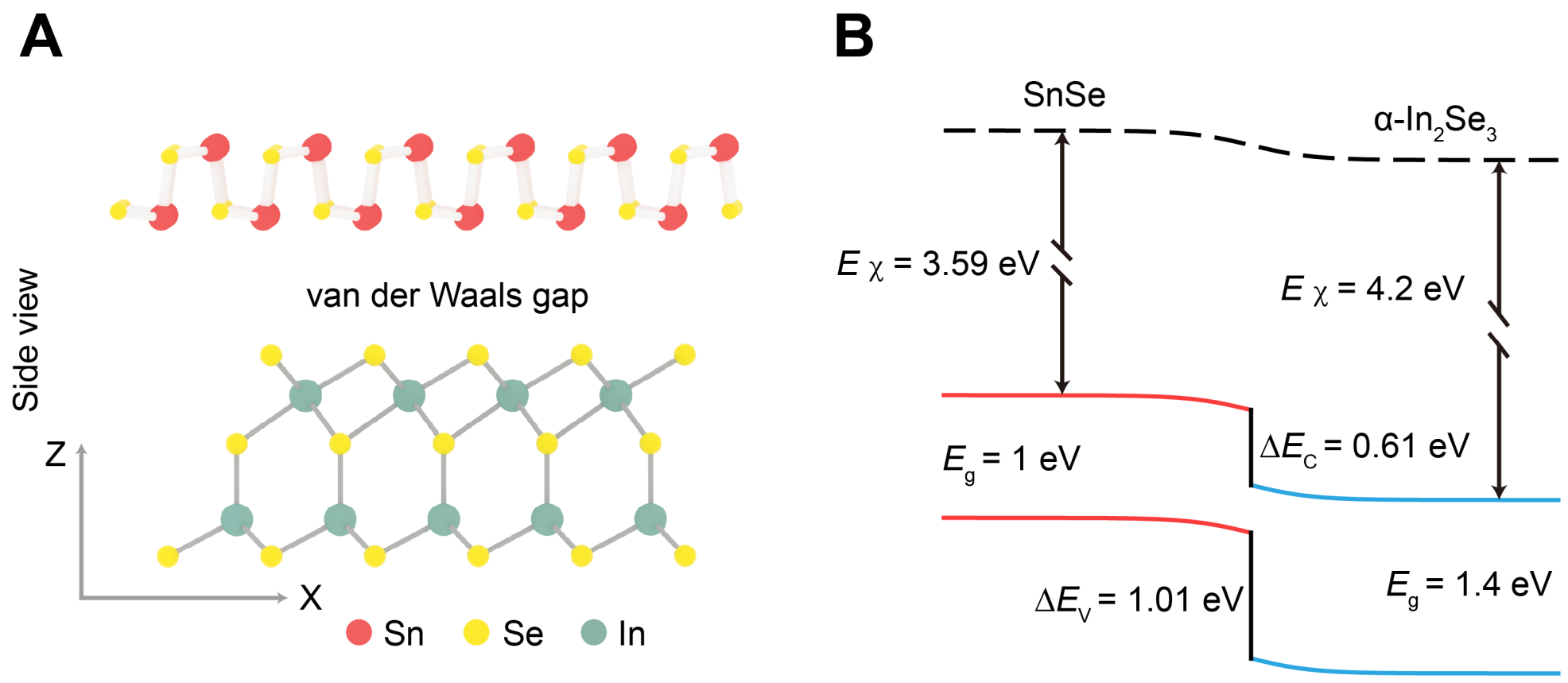

**Fig. S5. Crystal structures and band alignment of the SnSe/α-$In_2Se_3$ p-n heterojunction. (A)** Crystal structures of the SnSe (top) and α-$In_2Se_3$ (bottom). **(B)** Band alignment diagram of the SnSe/α-$In_2Se_3$ p-n junction in the spontaneous state.

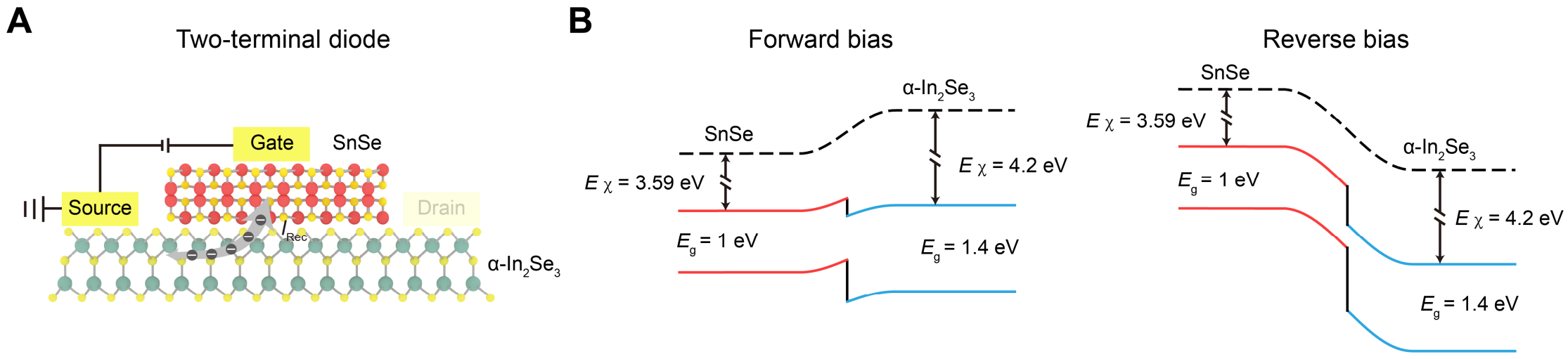

**Fig. S6. Two-terminal diode architecture and bias-dependent band alignments of the SnSe/α-$In_2Se_3$ p-n heterojunction. (A)** Schematic illustration of the two-terminal diode device. **(B)** Energy band diagrams of the heterojunction under forward and reverse bias.

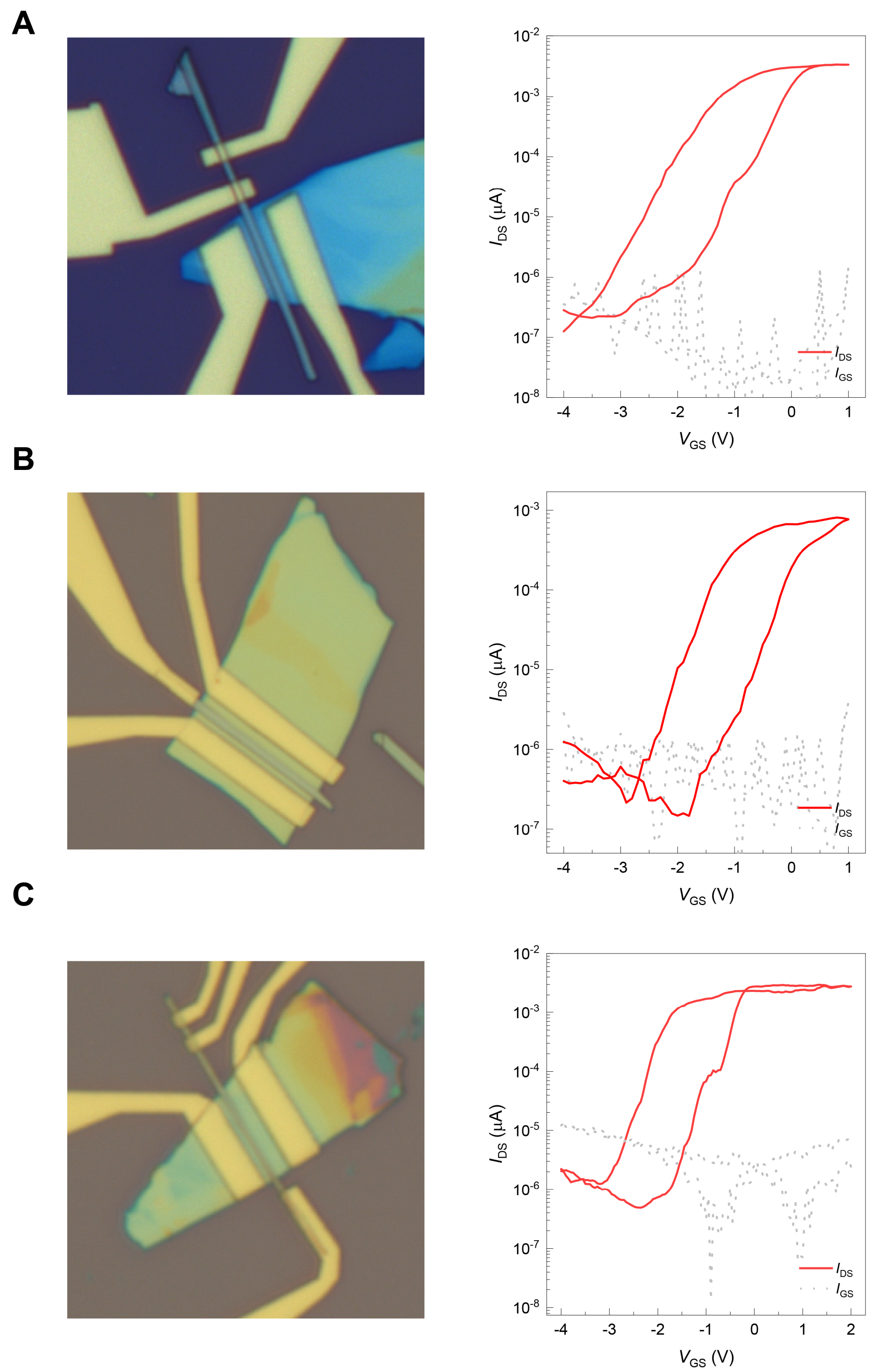

**Fig. S7. Transfer characteristics of other three representative Fe-JFET devices. (A)-(C) Left**: Optical images of these Fe-JFET devices. **Right**: Transfer characteristics of these Fe-JFET devices. Scale bar: 2 μm.

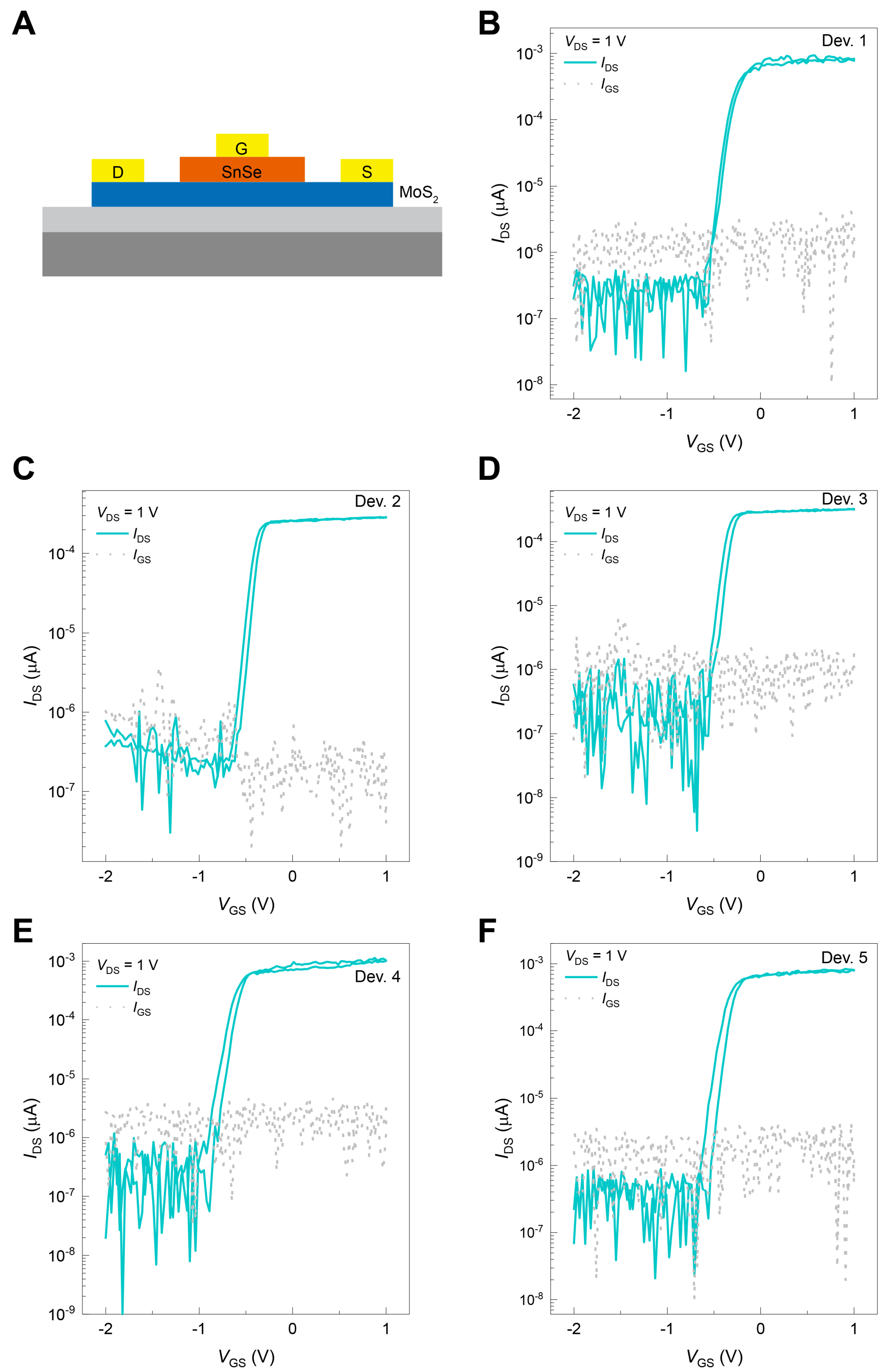

**Fig. S8. Transfer characteristics of JFETs with $MoS_2$ as the channel material. (A)** Schematic of the device structure, which is identical to the Fe-JFET. The difference lies that the ferro-channel is replaced with non-ferroelectric $MoS_2$. **(B)-(F)** Transfer curves of five devices (Dev.1 to Dev. 5), showing no obvious hysteresis loops, further confirming that the Fe-JFET's loop originate from the α-$In_2Se_3$ channel.

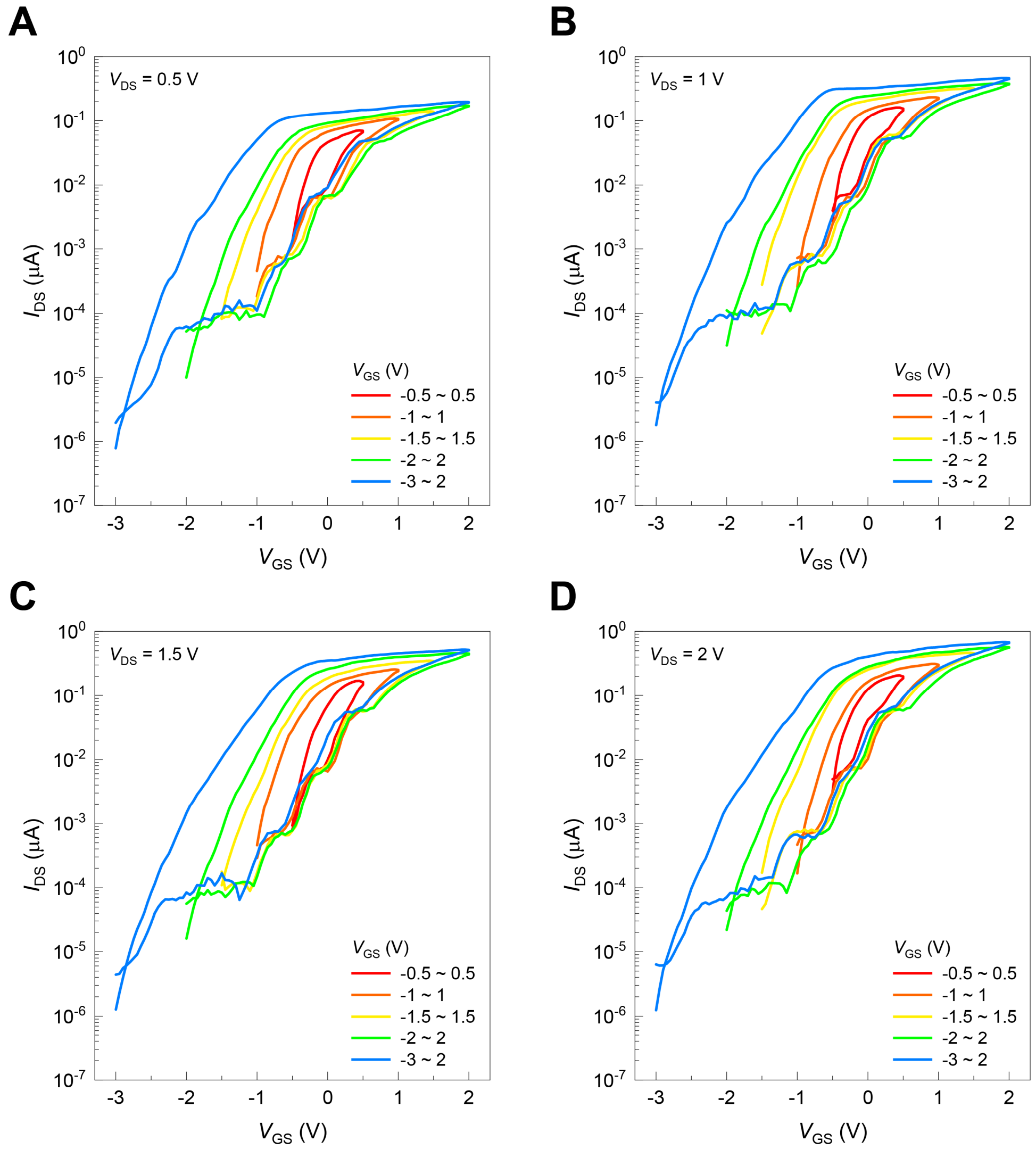

**Fig. S9. Fe-JFET transfer characteristics under different $V_{GS}$ sweeping ranges and $V_{DS}$. (A)** Transfer curves at $V_{DS}$ = 0.5 V, **(B)** Transfer curves at $V_{DS}$ = 1 V, **(C)** Transfer curves at $V_{DS}$ = 1.5 V and **(D)** Transfer curves at $V_{DS}$ = 0.5 V with $V_{GS}$ sweeping between -0.5 to 0.5 V, -1 to 1 V, -1.5 to 1.5 V, -2 to 2 V, and -3 to 2 V.

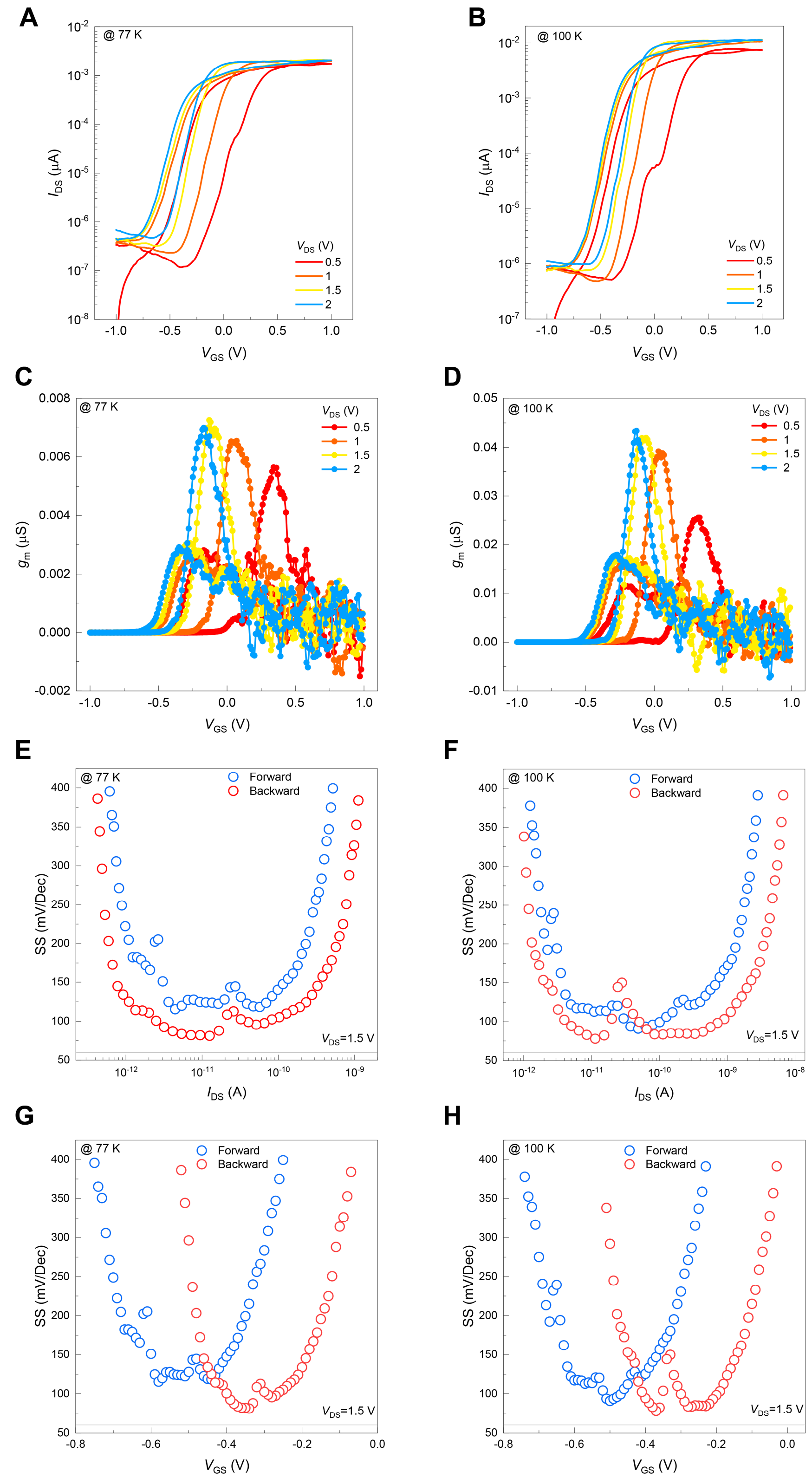

**Fig. S10. Fe-JFET electrical performance at low temperature. (A), (B)** Transfer curves with bidirectional sweeping at 77 K and 100 K, respectively. **(C), (D)** Corresponding $g_m$ versus $V_{GS}$ curve at 77 K and 100 K. **(E), (F)** Corresponding SS versus $I_{DS}$ curve at 77 K and 100 K. **(G), (H)** Corresponding SS versus $V_{GS}$ curve at 77 K and 100 K. The reduction in SS is attributed to the decrease in temperature.

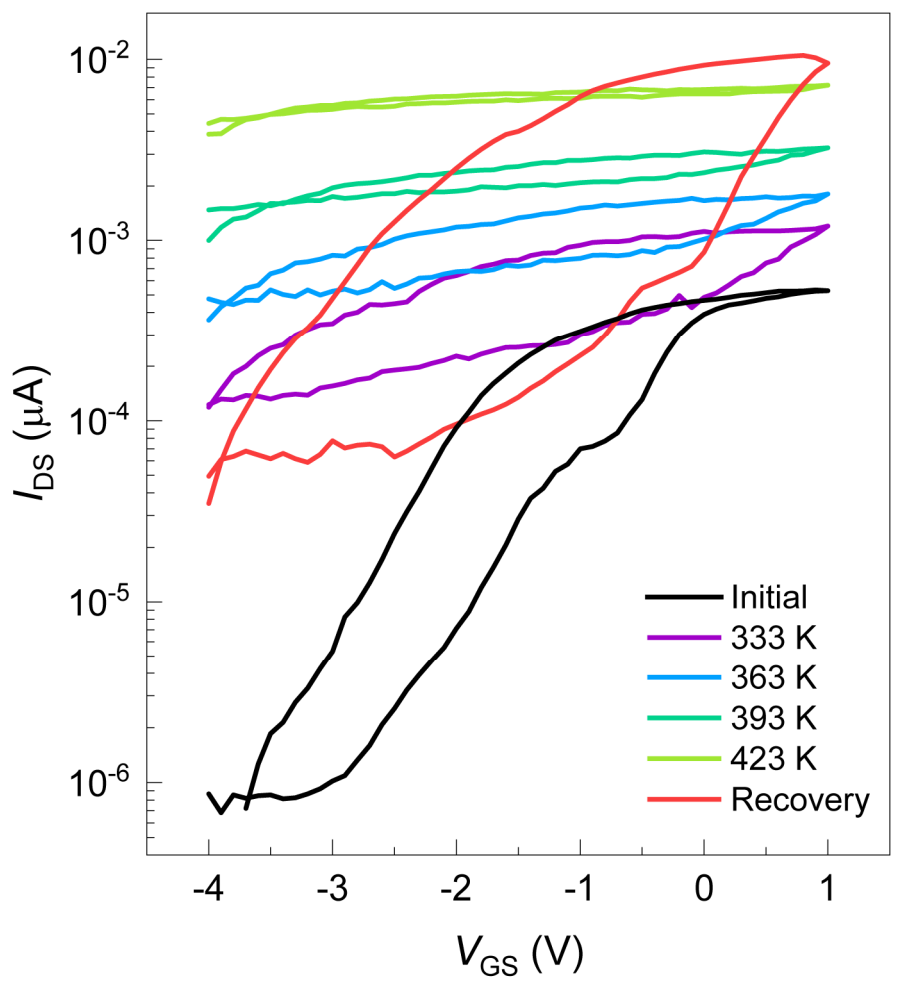

**Fig. S11. Fe-JFET temperature tolerance measurements.** The recovery of hysteresis window demonstrates the temperature robustness for our Fe-JFET.

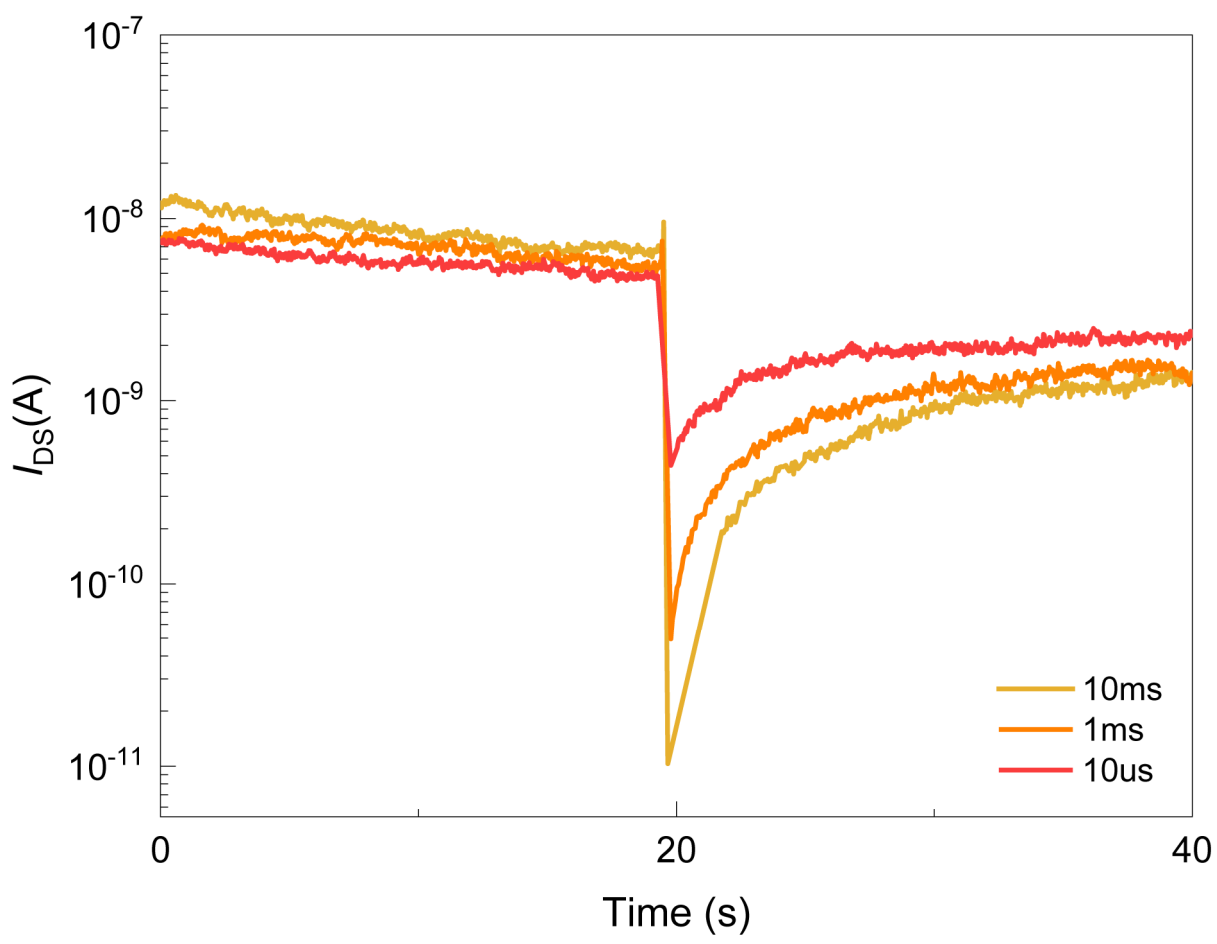

**Fig. S12. $I_{DS}$ transient variation under a 10 V pulse voltage with different widths.**

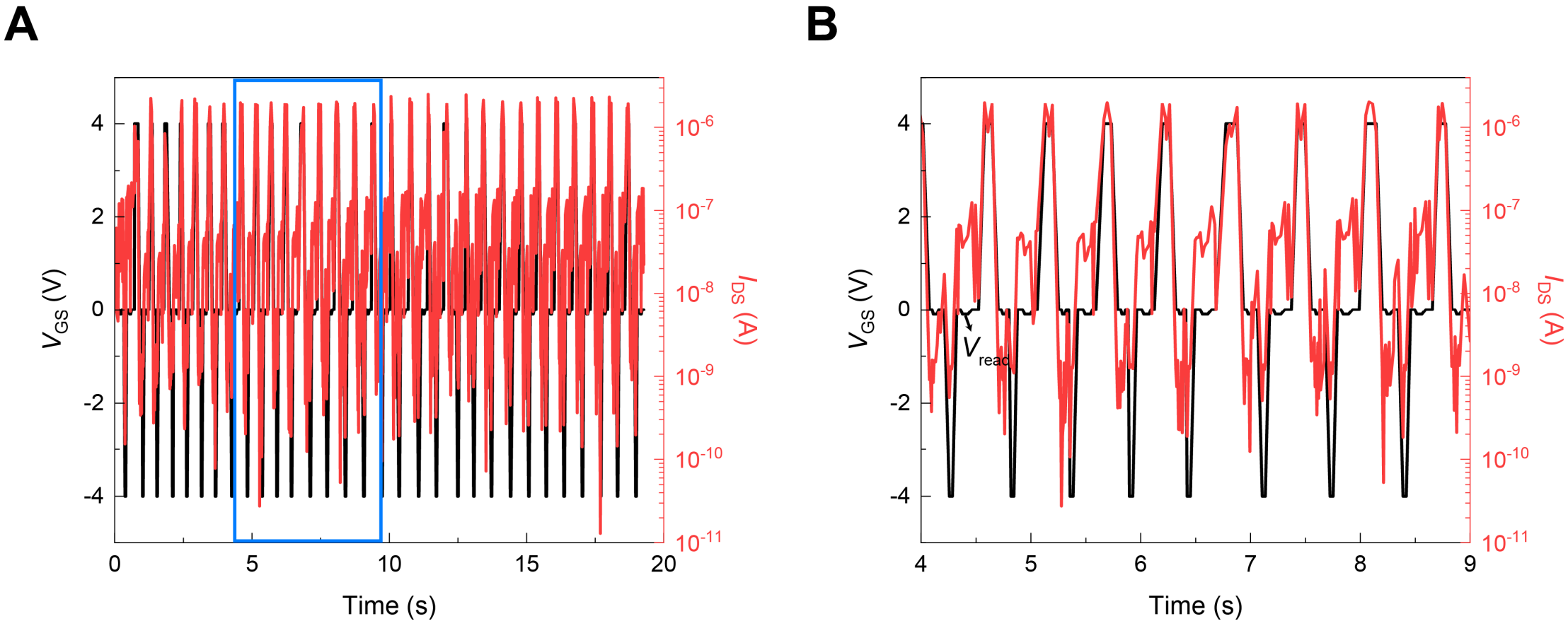

**Fig. S13. $V_{GS}$ waveform and $I_{DS}$ variation of the endurance tests. (A)** Partial segment of the endurance test. **(B)** Magnified view of the selected region in (A), illustrating the $I_{DS}$ evolution at the $V_{read}$ after each positive or negative pulsed $V_{GS}$ poling.

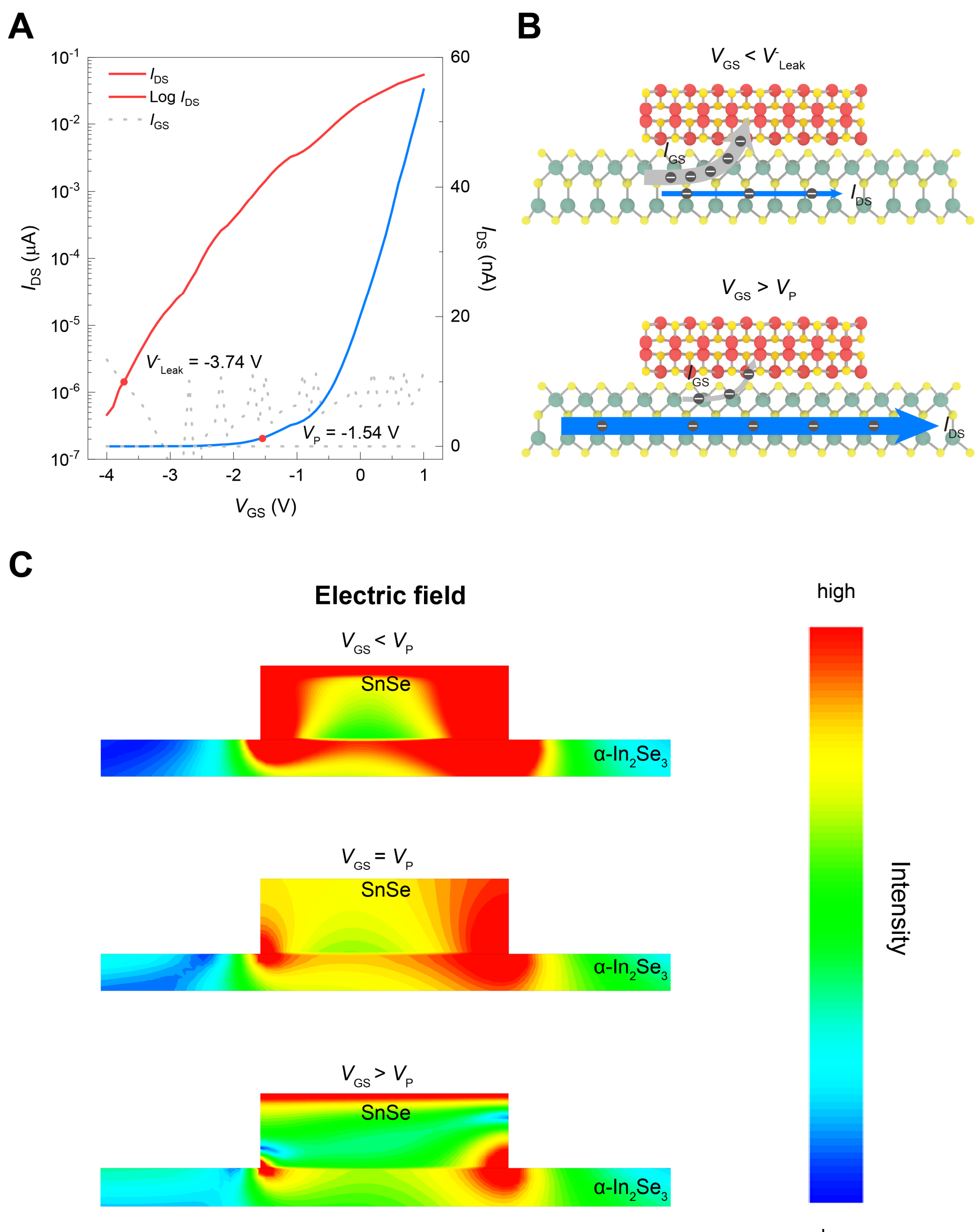

**Fig. S14. Mechanism analysis of the Fe-JFET transistor functionality. (A)** The linear and logarithmic transfer curve at $V_{DS}$ = 1 V. The threshold voltages for the leakage current ($V^-_{Leaky}$ = -3.74 V) and the pinch-off voltage ($V_P$ = -1.9 V) are labeled. **(B)** Schematic diagram illustrating the flow of channel current ($I_{DS}$) and leakage current ($I_{GS}$). **(C)** Electric field distribution simulated using TCAD under various $V_{GS}$.

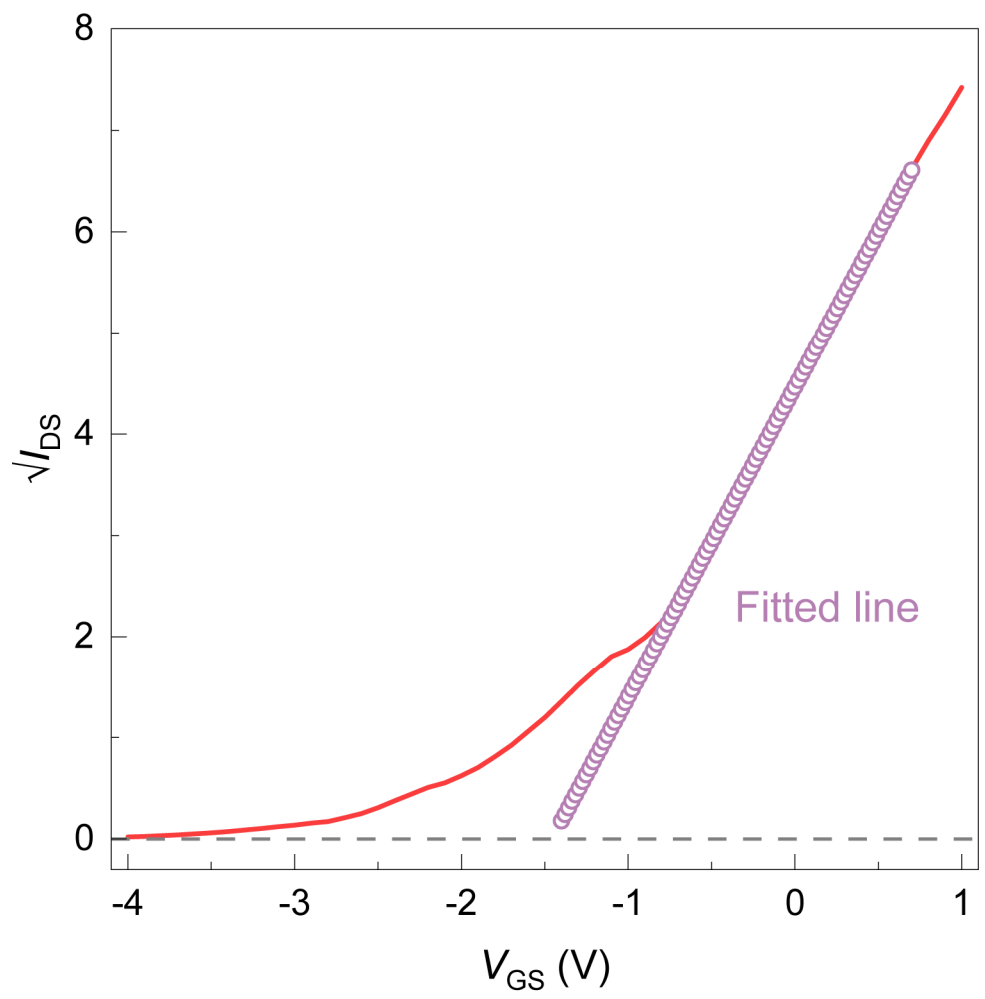

**Fig. S15.** $\sqrt{I_{DS}}$ **versus** $V_{GS}$ **curve of the Fe-JFET at** $V_{DS}$ **= 1 V.** The purple scatter line linearly fits the curve of $\sqrt{I_{DS}}$, and its intersection with x axis determines $V_P$ (-1.54 V).

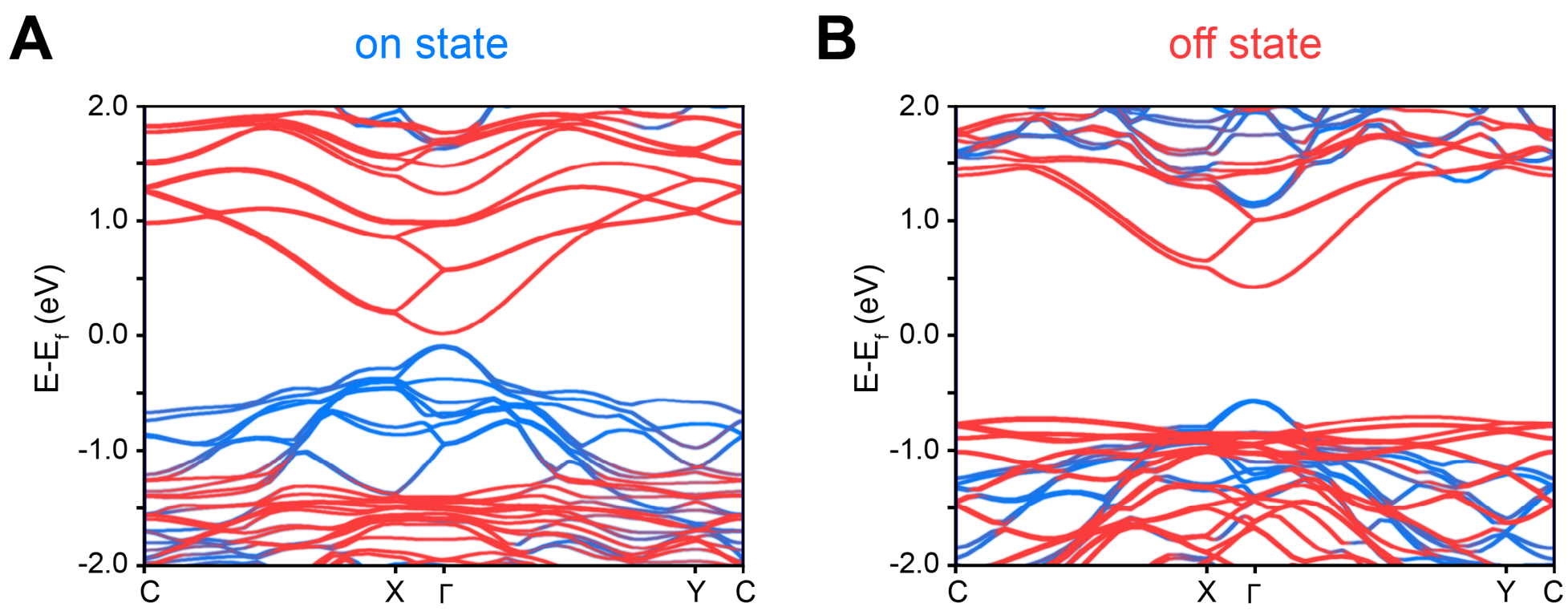

**Fig. S16. Variation of SnSe/α-$In_2Se_3$ p-n heterojunction energy band alignment induced by ferroelectric polarization charges.** Calculated energy band structures for **(A)** the on state and **(B)** the off state. The red and blue curves correspond to the projected contributions from α-$In_2Se_3$ and SnSe, respectively.

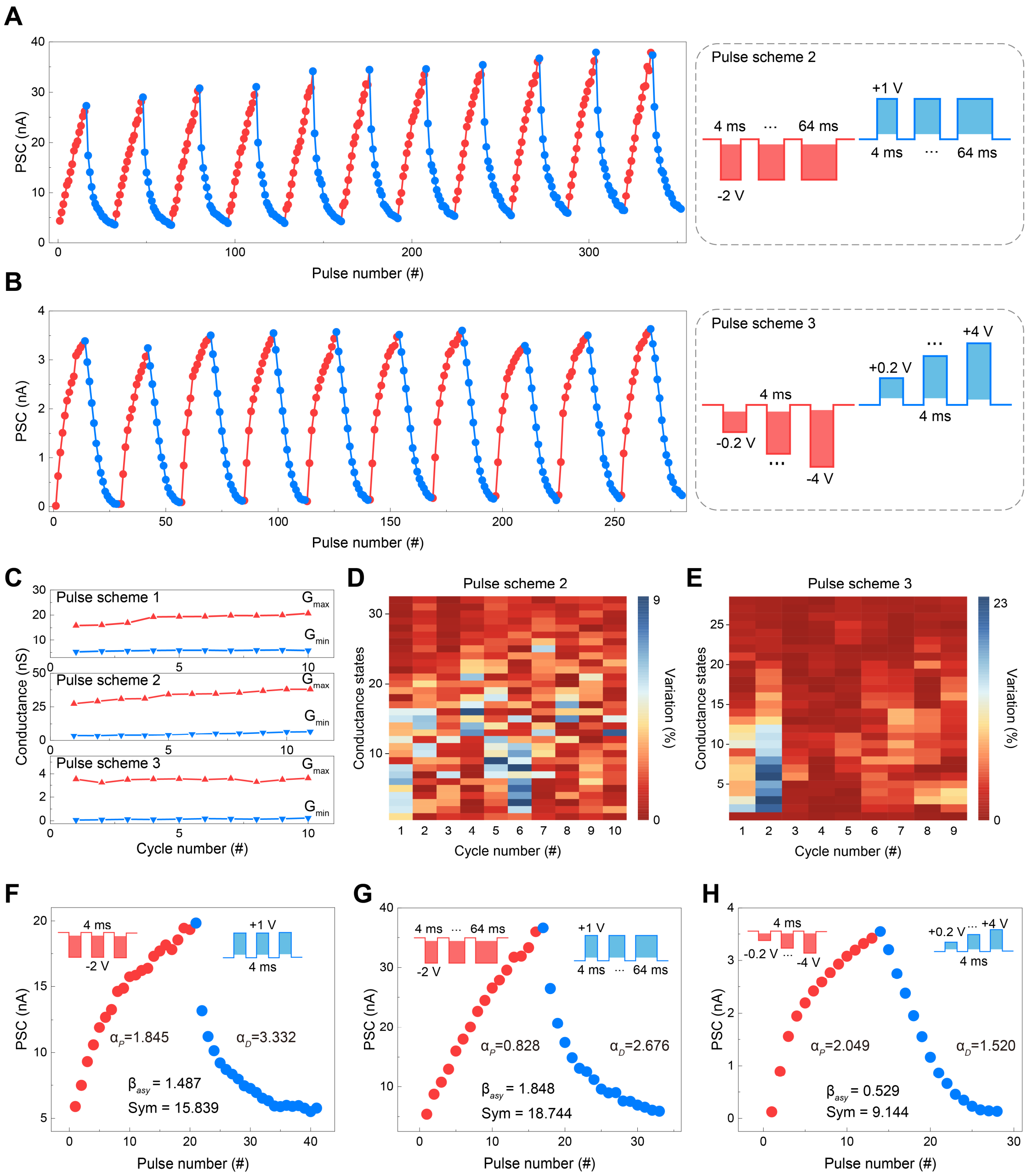

**Fig. S17. LTP and LTD characteristics under different pulse schemes. (A), (B)** LTP and LTD curves with the other two pulse schemes, including consecutive pulse voltages with constant amplitude and gradually increased width, as well as constant pulse width and gradually increased amplitude. **(C)** Extracted $G_{\max}$ and $G_{\min}$ from the LTP and LTD characteristics under three pulse schemes to exhibit conductance stability over 10 cycles. **(D), (E)** CCV heatmaps of relative variation across cycles for pulse schemes 2 and 3, respectively, indicating excellent stability. **(F)-(H)** Fitting of LTP and LTD PSC curves for pulse schemes 1, 2, and 3, highlighting key fitting parameters α, β and symmetry.

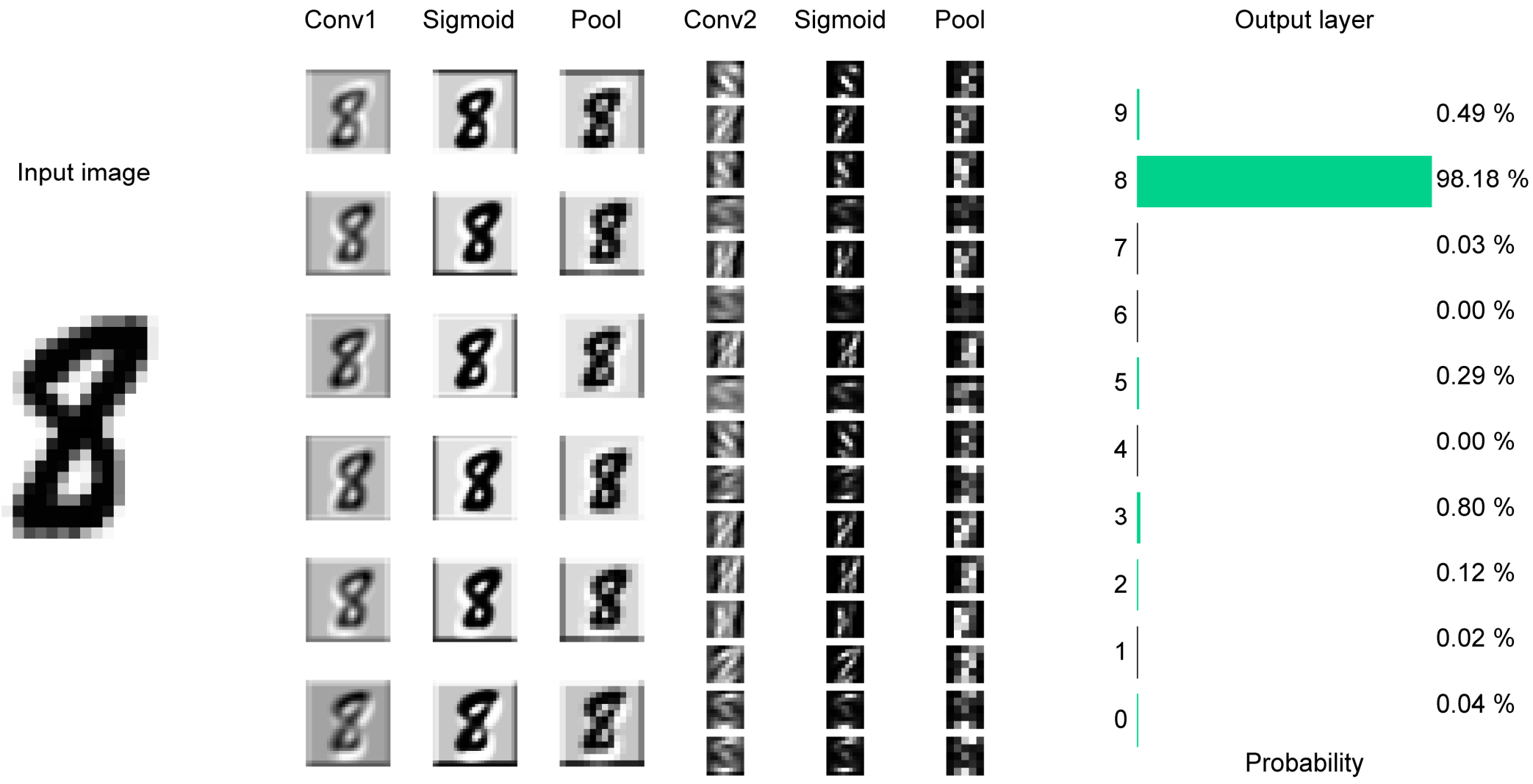

**Fig. S18. Step-by-step feature extraction and classification in our constructed CNN model.** The input image of the digit "8" undergoes sequential processing through two convolutional layers (Conv1 and Conv2), each followed by a Sigmoid activation and a pooling layer. The convolutional layers extract increasingly abstract features, while the pooling layers reduce spatial dimensions. The output layer displays the classification probabilities for each digit (0–9), with the model confidently predicting "8" at 98.18%, demonstrating high accurate recognition.